\documentclass[%
 aip, 
 pop, 
 amsmath,amssymb,
preprint,%
]{revtex4-1}

\usepackage{graphicx}
\usepackage{dcolumn}
\usepackage{bm}
\usepackage{cancel} 
\usepackage{floatrow}
\usepackage[caption=false]{subfig}
\usepackage[breaklinks=true,colorlinks=true,linkcolor=blue,urlcolor=blue,citecolor=blue]{hyperref}
\usepackage{xcolor}
\hypersetup{
	colorlinks   = true, 
	urlcolor     = red, 
	linkcolor    = blue, 
	citecolor   = blue 
}
\usepackage{hypernat}

\usepackage{epstopdf}

\begin{document}
	
	
\title{Limits of applicability of  the quasilinear approximation to the  electrostatic wave-plasma interaction} 
	
	
	
\author{Georgios Zacharegkas}
\email[]{gzachare@physics.auth.gr}
\affiliation{Aristotle University of Thessaloniki, 52124 Thessaloniki, Greece}

\author{Heinz Isliker }
\email[]{isliker@astro.auth.gr}
\affiliation{Aristotle University of Thessaloniki, 52124 Thessaloniki, Greece}

\author{Loukas Vlahos}
\email[]{vlahos@astro.auth.gr}
\affiliation{Aristotle University of Thessaloniki, 52124 Thessaloniki, Greece}
	
	
\date{\today}
	
\begin{abstract}
The limitation of the Quasilinear Theory (QLT) to describe the diffusion of electrons and ions in velocity space when interacting with a spectrum of large amplitude electrostatic Langmuir, Upper and Lower hybrid waves, is analyzed. 
We analytically and numerically estimate the threshold for the amplitude of the waves above which the QLT breaks down, using a test particle code.  
The evolution of the velocity distribution, the velocity-space diffusion coefficients, the driven current, and the heating of the particles 
are investigated, for the interaction with 
small and large amplitude electrostatic waves, 
i.e.\ in both regimes, there where QLT is valid and there where it clearly 
breaks down. 
\end{abstract}
	
\pacs {52.20.-j, 52.65.-y, 52.35.Ra}
	
\maketitle 

\section{Introduction}
\label{sec:1-Introduction}

The theory that describes the interaction between charged particles with a spectrum of low-amplitude waves in plasmas, is termed quasilinear theory (QLT) and has been widely studied and used in many applications. The theory for non-relativistic particles has been analyzed by \citet{AIPformalism}, and 
the one for relativistic particles by \citet{lerche68}. 

\citet{tao11} tested the theory of \citet{AIPformalism} for parallel propagation of Whistler modes, using a relativistic particle code, and found perfect agreement between the theoretical and their numerical diffusion coefficients for low-amplitude modes, i.e. in the quasilinear regime, thus validating the theory. 

Although QLT is really practical and relatively simple, its downside lies in its limited applicability. The linearization approach that QLT adopts can be applied only in low-amplitude turbulence, where nonlinear wave-particle interactions and nonlinear wave-wave couplings are negligible.

 QLT can also be applied when a single wave is assumed. The stochastic heating of ions by a single lower hybrid wave, propagating almost purely perpendicularly to a uniform magnetic field,  was studied by \citet{karney77II}, and the diffusion coefficients in velocity space were derived. \citet{karneyAIP78} proved that if the wave amplitude remains below a certain threshold, the diffusion coefficient is similar to the one  estimated by \citet{AIPformalism} and the damping of the wave is linear. Above this threshold though, and for finite wave amplitudes where nonlinear effects appear, stochasticity governs the phase-space, the resonances are broadened, and the ions get directly heated by the wave, irrespective of how close the frequency of the wave is to a cyclotron harmonic (i.e.\ the resonance condition is not a necessary condition for efficient particle heating). Detailed discussion on the nature and physical interpretation of this behavior can be found in several articles \cite{kaufman79, gell80, shklyar86}.  \citet{sen96} studied the role of an additional, second, obliquely propagating wave and  found that the particle motion in such cases is much more complicated, and that the stochasticity threshold for the first wave's amplitude is reduced due to nonlinear modification of the cyclotron resonances, caused by the presence of the second wave.

The investigation of the validity of QLT for a continuous spectrum of waves is a subject of interest as well, since it is a more realistic case. \citet{lange13} showed that for steep spectra, QLT is not able to predict a resonance (this point was also discussed by \citet{shalchi04}), since it manifests an irregularity around the resonant point. This is a known problem with QLT\cite{spanier12}.

Our analysis in this article is based on resonant wave-particle interactions that lead to acceleration and heating of the particles in the presence of electrostatic waves (Langmuir waves (LW), Upper Hybrid (UH), and Lower Hybrid (LH) waves). We focus on the theoretical description of the case of non-relativistic particles, and by using the test particle approach we estimate the validity of QLT by comparing the analytical and numerical diffusion coefficients and velocity distribution functions of the particles. We begin by setting the wave amplitude low enough for QLT to be valid, and then gradually increase the amplitudes of the waves and monitor the departure from the predictions of the QLT.  We start our analysis with the case of LW, which have extensively been discussed in the literature \cite{galeev69}, and we use this study as a basis to test our numerical model and then expand our study to the cases of UH and LH waves.

This paper is structured as follows. In Section~\ref{sec:2-general theory} a brief general description of our model is presented.  Sections~\ref{sec4:Langmuir section} and~\ref{sec5:hybrid} summarize the analytical and numerical results for the cases of Langmuir, UH, and LH waves, respectively. We then discuss the results in Section~\ref{sec:6-Conclusions}. Additionally, Appendix~\ref{sec:appendixA-dispersion relations} gives some details on the derivation of the UH and LH dispersion relations, and Appendix~\ref{sec:appendixB-diffusion coefficients} includes some calculations required for deriving the diffusion coefficients in these two cases.

\section{Model description}
\label{sec:2-general theory}


\subsection{Initial Settings}
\label{subsubsec:QLT equations}

We consider a non-relativistic and collisionless plasma of temperature $T$ and number density $n_0$, with total plasma thermal energy $\mathcal{W}_{tot}$, with the plasma being either magnetized or unmagnetized. The electric field $\mathbf{E}(\mathbf{x},t)$ is considered to be a superposition of many electrostatic modes $\delta \mathbf{E}_\mathbf{k}(\mathbf{x},t)$, each having a random phase $\vartheta_\mathbf{k}$ and energy density  \cite{dwight93, galeev69, kolm95}
\begin{equation}\label{eqintr:electric energy per component}
\mathcal{W}_\mathbf{k}(t) \equiv \frac{| \delta \mathbf{E}_\mathbf{k}|^2}{8\pi} = ak^{-5/3},
\end{equation}
i.e.\ we assume that the wave energy density spectrum follows a Kolmogorov power-law. The normalization constant, $a$, is such that it satisfies the condition
\begin{equation}\label{eqintr:wave energy density initialization}
W_0 = \int _{k_{\min}}^{k_{\max}} \mathcal{W}_\mathbf{k}(t=0) dk \equiv \kappa W_{\rm tot}
\end{equation}
at initial time $t=0$,
where $W_0$ is the total wave energy density, which is taken to be a fraction $\kappa$ of $W_{\rm tot}$, and $k_{min}, k_{max}$ are the limits of the wave spectrum. The coefficient  $ \kappa $ is a free parameter in our analysis.

We assume that the particles have initially a velocity distribution function which is of the form of an isotropic Maxwellian, 
\begin{equation}\label{eqintr:Initial Maxwellian}
f^s(\mathbf{v}, t=0)\equiv \frac{n_0}{\left( \sqrt{2\pi} \; u_{\rm th,s}\right) ^3}\exp \left( -\frac{\mathbf{v}^2}{2u_{\rm th,s}^2} \right) 
\end{equation}
where $u_{\rm th,s}\equiv (T_s/m_s)^{1/2}$ is their thermal speed, and $s=e/i$ denotes electrons/ions. 

The wave-particle interaction modifies the initial velocity distribution by absorbing energy from the waves.  
The QLT is built on the assumption that we can split the distribution function into two parts, namely the averaged slowly varying part, $f^s_0(\mathbf{v},t) \equiv \langle f^s(\mathbf{x},\mathbf{v},t) \rangle _\mathbf{x}$, and the first order rapidly varying perturbation, $\delta f^s(\mathbf{x},\mathbf{v},t)\equiv f^s(\mathbf{x},\mathbf{v},t)-\langle f^s(\mathbf{x},\mathbf{v},t) \rangle _\mathbf{x}$. 
In this context then, one can derive the diffusion equation for $f^s_0$ by solving the set of equations consisting of the averaged and linearized Vlasov equations, combined with the Maxwell equations. This diffusion equation can be written as
\begin{equation}\label{eqth:diff eq}
\frac{\partial f^s_0(\mathbf{v},t)}{\partial t}=\nabla _{\mathbf{v}} \cdot \left[ D_s^{\rm QL} (\mathbf{v},t) \nabla_{\mathbf{v}}f^s_0(\mathbf{v},t) \right]
\end{equation}
with the velocity diffusion tensor \cite{AIPformalism}
\begin{equation}\label{eqth:velocity diff tensor}
	D_s^{\rm QL}(\mathbf{v},t)\equiv i \frac{e^2}{m_s^2} \int d^3\mathbf{k} \sum \limits _{n=-\infty}^\infty \frac{\left( \mathbf{d}^s_{n,\mathbf{k}}\right) ^* \mathbf{d}^s_{n,\mathbf{k}}}{\sigma _\mathbf{k}-k_\| v_\| -n\Omega _s},
\end{equation}
where the vector
\begin{eqnarray}\label{eqth:velocity diff vector}
	\mathbf{d}^s_{n,\mathbf{k}} \equiv \mathcal{E}^s_{n,\mathbf{k}} && \left[ \mathbf{\hat{e}}_\perp \left( 1-\frac{k_\| v_\| }{\sigma _\mathbf{k}} \right) +\hat{\mathbf{e}}_\| \frac{k_\| v_\perp }{\sigma _\mathbf{k}} \right] +\left( E_\| \right)_\mathbf{k} J_n (\zeta _\mathbf{k}^s) \nonumber \\
	&&\times \left[ \mathbf{\hat{e}}_\| \left( 1-\frac{n\Omega _s}{\sigma _\mathbf{k}} \right) + \mathbf{\hat{e}}_\perp \frac{n\Omega _s}{\sigma _\mathbf{k}} \frac{v_\| }{v_\perp } \right] ,
\end{eqnarray}
with $\mathbf{\hat{e}}_\perp \equiv \mathbf{v}_\perp /|\mathbf{v}_\perp|$ and $\mathbf{\hat{e}}_\| \equiv \mathbf{\hat{e}}_\mathbf{z}$,
by assuming that the magnetic field points along the $z$-direction, and where $\sigma_\mathbf{k} = \omega_\mathbf{k}+i\gamma_\mathbf{k}$ is the complex frequency of the waves. Here, $J_n(\zeta _\mathbf{k}^s)$ are the Bessel functions of first kind, where $\zeta _\mathbf{k}^s \equiv k_\perp v_\perp /\Omega_s$, $\Omega_s$ is the gyrofrequency  and the symbols $\|$ and $\perp$ denote the direction parallel and perpendicular to the magnetic field, respectively. Finally, also the definition
\begin{eqnarray}\label{eqth:convenience function Epsilon}
	\mathcal{E}^s_{n,\mathbf{k}} & \equiv & \frac{E_x (\mathbf{k})}{2}\left[ e^{i\psi}J_{n+1}(\zeta _\mathbf{k}^s)+e^{-i\psi}J_{n-1}(\zeta _\mathbf{k}^s) \right] \nonumber \\
	&& +i\frac{E_y (\mathbf{k})}{2}\left[ e^{i\psi}J_{n+1}(\zeta _\mathbf{k}^s)-e^{-i\psi}J_{n-1}(\zeta _\mathbf{k}^s) \right]
\end{eqnarray}
is used, where $\psi$ is the polar angle of $\mathbf{k}_\perp$.

In order to derive the diffusion rates for specific kinds of waves, we must couple Eq.~(\ref{eqth:diff eq}) with the linearized Poisson equation,
\begin{equation}\label{eqth:Poisson equation linear}
i \; \mathbf{k} \cdot \delta \mathbf{E}_\mathbf{k}(t)=4\pi e\int d^3\mathbf{v} \; \left[ \delta f^i_\mathbf{k}(\mathbf{v},t) - \delta f^e_\mathbf{k}(\mathbf{v},t) \right]  ,
\end{equation}
from which the linear dispersion relation can be derived. The  $\delta f^s_\mathbf{k}$ that is needed can be expressed as \cite{AIPformalism}
\begin{eqnarray}\label{eqth:deltaf analytic solution}
	\delta f^s_{\mathbf{k}} = -i \frac{(\pm e)}{m_s}E_\mathbf{k} \sum \limits_{n,m=-\infty}^{\infty} && \frac{nJ_n(\zeta _\mathbf{k}^s)J_m(\zeta _\mathbf{k}^s)}{\zeta_\mathbf{k}^s} \nonumber \\
	&& \times \frac{e^{i(m-n)(\phi-\psi)}}{\sigma_\mathbf{k} -n\Omega_s} \frac{\partial f^s_0}{\partial v_\perp}  ,
\end{eqnarray}
and the derivation of the dispersion relation then is straightforward (as briefly presented in Appendix~\ref{sec:appendixA-dispersion relations}).

The QLT describes the slowly varying distribution function $f_0^s$, 
and in order to be valid, it needs to be ensured that $f_0^s$ changes slowly enough, and more specifically its diffusive relaxation time, $\tau_R$ (defined below), must be clearly longer than the wave-particle interaction time-scale \cite{chirikov79, itoh09, shapiro97} $|\gamma_\mathbf{k}|^{-1}$
(with $\gamma_\mathbf{k}$ the growth rate, see below), thus
\begin{equation}\label{eqth:slow damping condition}
\tau_R \gg |\gamma_\mathbf{k}|^{-1}.
\end{equation}
If that is the case, the space-averaged distribution function changes slowly enough, so that particles, which gyrate around the magnetic field and diffuse in velocity-space, do not experience any changes of $f^s_0$ on their characteristic gyro-motion and wave-motion timescales.\cite{AIPformalism,chew56} Then, any dependence of $f^s_0$ on the velocity polar angle, $\phi =\tan^{-1} \left( v_y/v_x \right)$, is weak and averages out over a complete rotation from $0$ to $2\pi$, and therefore the diffusion process becomes two-dimensional\cite{AIPformalism} 
in the ($v_\perp,v_\|$)-space.

In all the three normal modes (LW, UH, LH)  that we will study, we will derive the upper limit for the wave amplitude for the QLT's applicability, using condition (\ref{eqth:slow damping condition}). 

  When this condition is not satisfied, the rapid energy exchange during the wave-particle interaction, and hence the rapid distribution function modification, does not allow the application of QLT's equations in the form presented above. 
  In addition, for QLT to be applicable in each of these cases, the Chirikov overlap criterion must be satisfied, so that the particles do not get trapped inside enhanced electric field structures, but travel relatively unhindered in velocity space. In more specific terms, this condition can be expressed \cite{itoh09} as $1/\tau_b<1/\tau_{ac}$, where $\tau_b$ is the bounce time-scale of the particles in the field, and $\tau_{ac}$ is the electric field's autocorrelation time-scale.

\subsection{Numerical model}
\label{subsec:numerical model}

In our numerical model, we use a large  a number of test particles, initially randomly distributed in space inside a periodic and cubic box in the plasma, and obeying a Maxwellian distribution function (\ref{eqintr:Initial Maxwellian}) in velocity-space. The  spectrum of waves used has $k\in [k_{\min},k_{\max}]$, and in total it 
initially carries the energy density $W_0$. 

A test particle evolves according to the Lorentz force, $d\mathbf{p} /dt=\pm  e\left( \mathbf{E}+\boldsymbol{\beta} \times \mathbf{B} \right)$, where $\boldsymbol{\beta} \equiv \mathbf{v}/c$, $\mathbf{v}$ is the velocity
and $\mathbf{p}$ the momentum. With $\mathbf{p}=\gamma m_s \mathbf{v}$, where the relativistic factor is $\gamma  \equiv \left( 1-\boldsymbol{\beta}^2\right) ^{-1/2}$, we can write the equations of motion in the following form,
\begin{equation}\label{eqnum:Lorentz eq of motion}
\frac{d\boldsymbol{\beta}}{dt} = \mp \frac{e}{c \gamma m_s}(\boldsymbol{\beta} \cdot \mathbf{E})\boldsymbol{\beta} \pm \frac{e}{c\gamma m_s}\mathbf{E}+\Omega_s \boldsymbol{\beta} \times \hat{\mathbf{e}}_\|
\end{equation}
in which $\Omega_s$ is the relativistic gyrofrequency, and where we also used the relation $d(c^2\gamma m_s)/dt = \pm e\mathbf{v}\cdot \mathbf{E}$.\cite{bitten04}.

Solving  the equations of motion, the numerical diffusion coefficient is calculated according to 
\begin{equation}\label{eqnum:diff coeff}
D \equiv \lim \limits_{\delta t\rightarrow 0} \frac{\langle \delta \mathbf{v} ^2 \rangle}{2 \delta t},
\end{equation}
in which the average is taken over the particles, divided into groups of similar initial velocities, and also $\delta \mathbf{v}=\mathbf{v}(t)-\mathbf{v}(t-\delta t)$, with relatively small values for $\delta t$, as indicated in the applications below \cite{Ragwitz2001}.

In the simulations we consider times such that $|\gamma_k|t\ll 1$, 
during which the energy loss of the waves is transferred to the test particles. Specifically, in the case of Langmuir waves, $\gamma_\mathbf{k} \propto (\partial f/\partial v)_{v\simeq \omega_e/k}$, and the formation of a plateau in the resonant region $v\simeq \omega_k/k$ quickly leads $\gamma_k$ to vanish. In the case of UH waves, the magnetic field is taken to be strong so $|\gamma_k|$ is  small compared to $|\Omega_e|$, and remains unchanged for the resonance condition $\omega_k\simeq |\Omega_e|$. The same is true in the case of LH waves, but here the resonance condition is $\omega_\mathbf{k}\simeq \Omega_i$ (see Section~\ref{sec5:hybrid}, and Appendix~\ref{sec:appendixA-dispersion relations}). Therefore, $\mathcal{W}_k(t)= \mathcal{W}_k(0)\exp (2\gamma_kt)\simeq \mathcal{W}_k(0) (1+2\gamma_kt) \simeq \mathcal{W}_\mathbf{k}(0)$, hence the wave energy changes only slightly during our simulations.

\section{Langmuir Waves}
\label{sec4:Langmuir section}

In this section we perform a first check of the QLT prediction about the maximum wave energy limit beyond which it stops being valid for the LW. The theory is presented briefly, since it is extensively analyzed in the literature (see for example \citet{galeev69}), and it is then tested using the analytical formulas
and the numerical estimates.

\subsection{Analytical predictions}
\label{subsec:langmuir theory}

In the absence of a magnetic field, $\Omega_e=0$, we can select $\hat{e}_\mathbf{z}$ as the propagation direction for the waves. Then, for $\mathbf{E}_\mathbf{k}=E_\mathbf{k}\mathbf{\hat{e}}_\mathbf{z}$, from Eq.~(\ref{eqth:convenience function Epsilon}) it is easy to show that $\mathcal{E}_{n,\mathbf{k}}=0$. Taking first the purely parallel propagation limit, $k_\perp \rightarrow 0^+$ and then the zero-magnetic field limit, $\Omega_e\rightarrow 0^-$, the Bessel function in $\mathbf{d}_{n,\mathbf{k}}$ is replaced by unity for $n=0$ in the sum of Eq.~(\ref{eqth:velocity diff tensor}), since $J_n(0)=\delta_{n0}$. Thus, the diffusion tensor in Eq.~(\ref{eqth:velocity diff tensor}) has only one non-zero component, the $\mathbf{\hat{e}}_\mathbf{z}\mathbf{\hat{e}}_\mathbf{z}$ term, which is simplified to
\begin{widetext}
\begin{equation}\label{eqthLM:diff coeff initial form}
	D^{\rm QL}(v) = 8\pi^2 \left( \frac{e}{m_e} \right)^2 \int d^3\mathbf{k} \left[ \mathcal{P} \frac{\gamma_\mathbf{k} \mathcal{W}_\mathbf{k}}{\left( \omega_\mathbf{k}-kv \right)^2+\gamma_k^2} + \pi \mathcal{W}_\mathbf{k} \delta \left( \omega_\mathbf{k}-kv \right) \right] ,
\end{equation}
\end{widetext}
after dropping every unnecessary index, since  the diffusion is one-dimensional and only electrons are considered, and where the symbol $\mathcal{P}$ denotes principal value. The imaginary part of this sum does not contribute, since it vanishes in the summation. Eq.~(\ref{eqthLM:diff coeff initial form}) is the already known \cite{galeev69} diffusion coefficient for Langmuir oscillations which takes into consideration the resonant particles through the delta function selection rule, as well as the non-resonant particles, which form the bulk distribution, through the integral over the principal value. Resonances involve particles with velocities $v\simeq \omega_\mathbf{k}/k$, which resonate with the corresponding waves.

Solving the linearized Vlasov equation for $\delta f_\mathbf{k}$ and substituting the solution into Eq.~(\ref{eqth:Poisson equation linear}) (in which only the electrons will appear due to the high frequency of the oscillations), it can be shown that the real part of the resulting equation has the solution $\omega_\mathbf{k}^2=\omega_e^2 \left[ 1 + (3/2) \left( k\lambda_e \right)^2 \right]$, where $\lambda_e$ is the electron Debye length, and which is expected since it expresses the Langmuir wave dispersion relation. 
Concerning the imaginary part $\gamma_\mathbf{k}$ of the frequency, we assume that $|\gamma_\mathbf{k}|\ll \omega_\mathbf{k}$, and from the resulting imaginary part of Eq.~(\ref{eqth:Poisson equation linear}), combined with the previous solution for $\omega_\mathbf{k}$, we get the solution for $\gamma_\mathbf{k}$, which is
\begin{equation}\label{eqthLM:imaginary frequency}
\gamma_\mathbf{k}\simeq \frac{\pi }{2}\frac{\omega_\mathbf{k}\omega_e^2}{n_0k^2}\left( \frac{\partial f_0}{\partial v} \right) _{v\simeq \omega_\mathbf{k}/k}.
\end{equation}
For a Maxwellian initial distribution, $f_0$, the damping coefficient is $\gamma_\mathbf{k} \propto -\exp \left[ - 2^{-1} \left( k\lambda_e \right) ^{-2}(\omega_\mathbf{k}/\omega_e )^2 \right]$, so the damping is weak if $k\lambda_e\ll 1$, in which case the expressions for the two frequency parts simplify to $\omega_\mathbf{k}\simeq \omega_e$ and $\gamma_\mathbf{k}\simeq (\pi/2)(\omega_e^3/n_0k^2)(\partial f_0/\partial v)_{v\simeq \omega_\mathbf{k}/k}$.

In the steady state limit, $\partial f(v,t\rightarrow \infty )/\partial t =0$, and in the resonance region the diffusion equation, Eq.~(\ref{eqth:diff eq}), implies that
\begin{equation}\label{eqthLM:quasi-stable condition}
\left. \left\{ \left[ \frac{\mathcal{W}_\mathbf{k}(t\rightarrow \infty)}{v} \right] \frac{\partial f_0(v,t\rightarrow \infty)}{\partial v} \right\} \right| _{v\simeq \omega_\mathbf{k}/k}  =0
\end{equation}
will hold. If we assume that the waves initially contain enough energy to fuel the whole energy exchange process (which is true in our case), then the only possible result of Eq.~(\ref{eqthLM:quasi-stable condition}) is the modification of the resonant distribution part, $f_0^{\rm r}$, as $[\partial f_0^{\rm r}(v,t\rightarrow \infty )/\partial v]_{v\simeq \omega_\mathbf{k}/k}=0$, i.e.\ a plateau will be formed and 
the resonant part of the distribution stops evolving. The plateau can easily be calculated as the mean value of $f_0$ in the resonant region in the time-asymptotic limit, and it is\cite{galeev69}
\begin{equation}\label{eqthLM:plateau}
\bar{f}_0^{\rm r}(t\rightarrow \infty) = \frac{1}{v_{\max} - v_{\min}} \int _{v_{\min}}^{v_{\max}} f(v,t\rightarrow \infty) dv ,
\end{equation}
where $\bar{f}_0^r(t\rightarrow \infty)$ is the 
value of the distribution at the plateau. The non-resonant distribution part, $f_0^{\rm nr}$, on the other hand, remains relatively unchanged, with a small increase in temperature\cite{galeev69}.


For a wave-spectrum of width $\Delta (\omega_\mathbf{k}/k)$, we can derive an approximate expression for the relaxation time $\tau_R=[\Delta  (\omega_\mathbf{k}/k)]^2/D^{\rm QL}$ of the particle diffusion as
\begin{equation}
	\tau_R \simeq \frac{n_0m_ek\left[ \Delta (\omega_e/k) \right]^3}{2\pi\omega_e^2W_0}.
\end{equation}
It then follows that in order for condition (\ref{eqth:slow damping condition}) to hold, $\kappa$ must be in the range indicated by
\begin{equation}\label{eqthLM:kappa restriction order}
	\kappa \ll \kappa_{\rm LM}^{\rm QL} \equiv \frac{n_0m_e\omega_e^2\left[ \Delta (\omega_e/k) \right]^3}{4 \sqrt{2\pi} k^2 u_{\rm th,e}^3 W_{\rm tot}} \exp  \left[ -\frac{1}{2} \left( \frac{\omega_e}{k u_{\rm th,e}} \right)^2 \right]
\end{equation}
(as an order of magnitude approximation), where the notation $\kappa _{\rm LM}^{\rm QL}$ is used to denote the upper limit of the range of $\kappa$ values for which QLT is valid, according to the analytical results. For a more accurate approximation of the upper limit for QLT's validity, we need to study the problem numerically.

\subsection{Numerical results}
\label{subsec:LM numerical results}

The parameters used for our numerical calculations are the total number of test-particles $N_p= 2 \times 10^4$, the density $n_0(cm^{-3})=10^9,$ the temperature $T(eV)=100$, and the wave phase velocity range $v_{min}/ u_{th}=2, u_{max}/u_{th}=4$. The initial velocity distribution of the test particles is a Maxwellian, as in Eq.~(\ref{eqintr:Initial Maxwellian}), and in space the particles are randomly distributed 
 in a box of linear size $2\cdot 10^5\lambda_e$. 
We consider a spectrum of $100$ waves, each assumed to have a random phase 
in $[0,2\pi]$, and their amplitudes follow the power-law of Eq.\ (\ref{eqintr:electric energy per component}). 

\begin{figure}[ht]
	\sidesubfloat[]{\includegraphics[width=0.4\columnwidth]{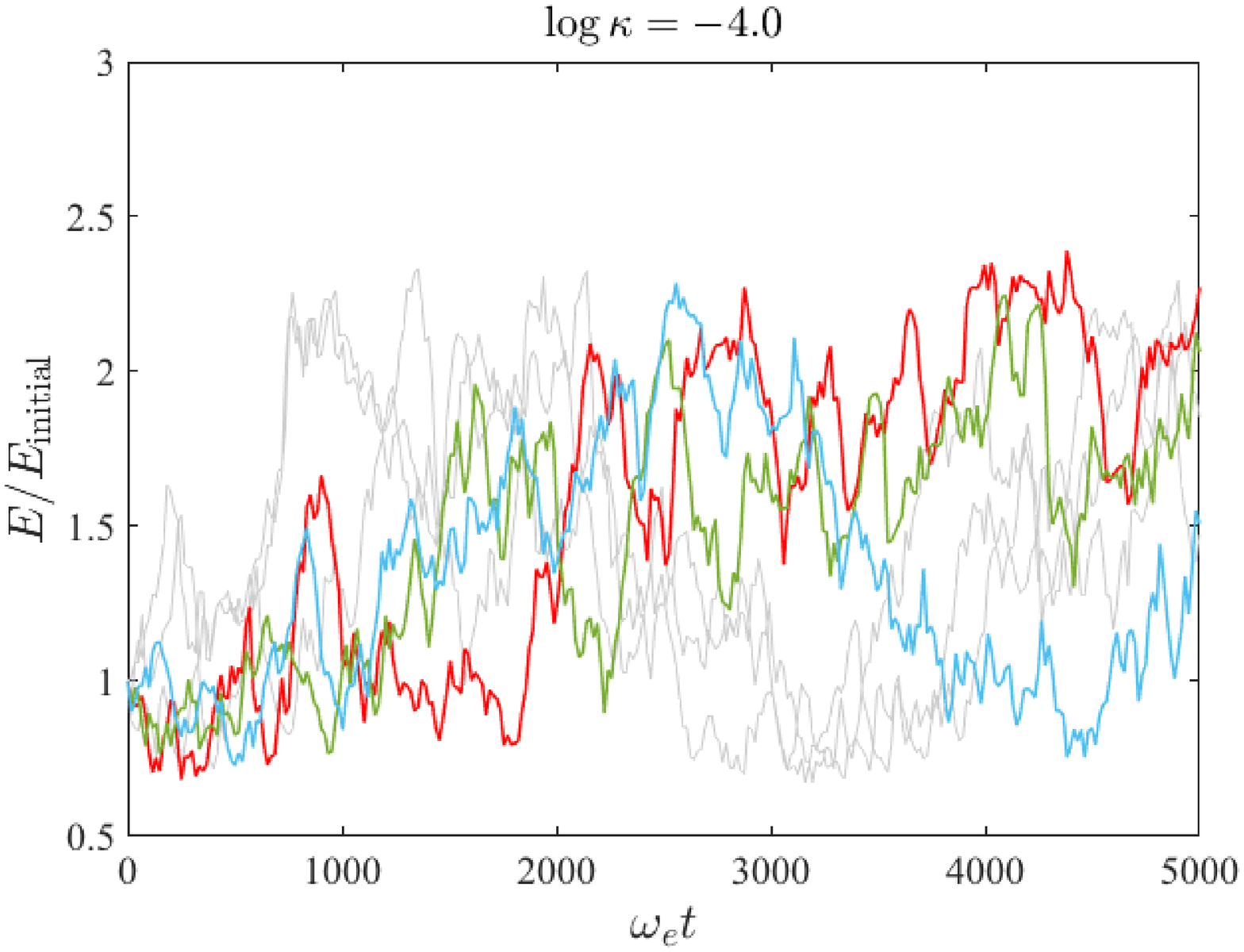}%
		\label{fig:langmuir energies -4}}\hfill
	\sidesubfloat[]{\includegraphics[width=0.4\columnwidth]{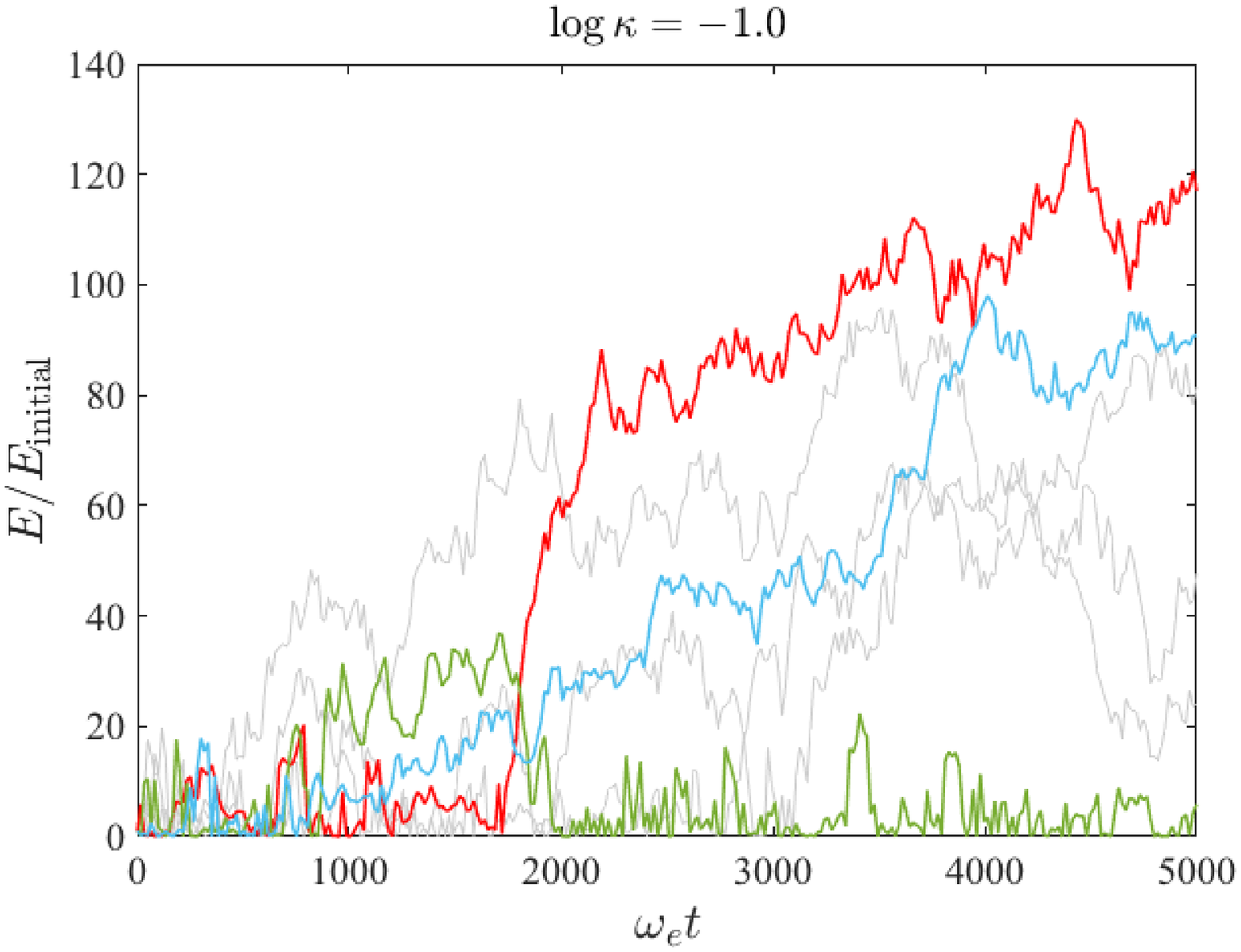}%
		\label{fig:langmuir energies -1}}\hfill
	\sidesubfloat[]{\includegraphics[width=0.4\columnwidth]{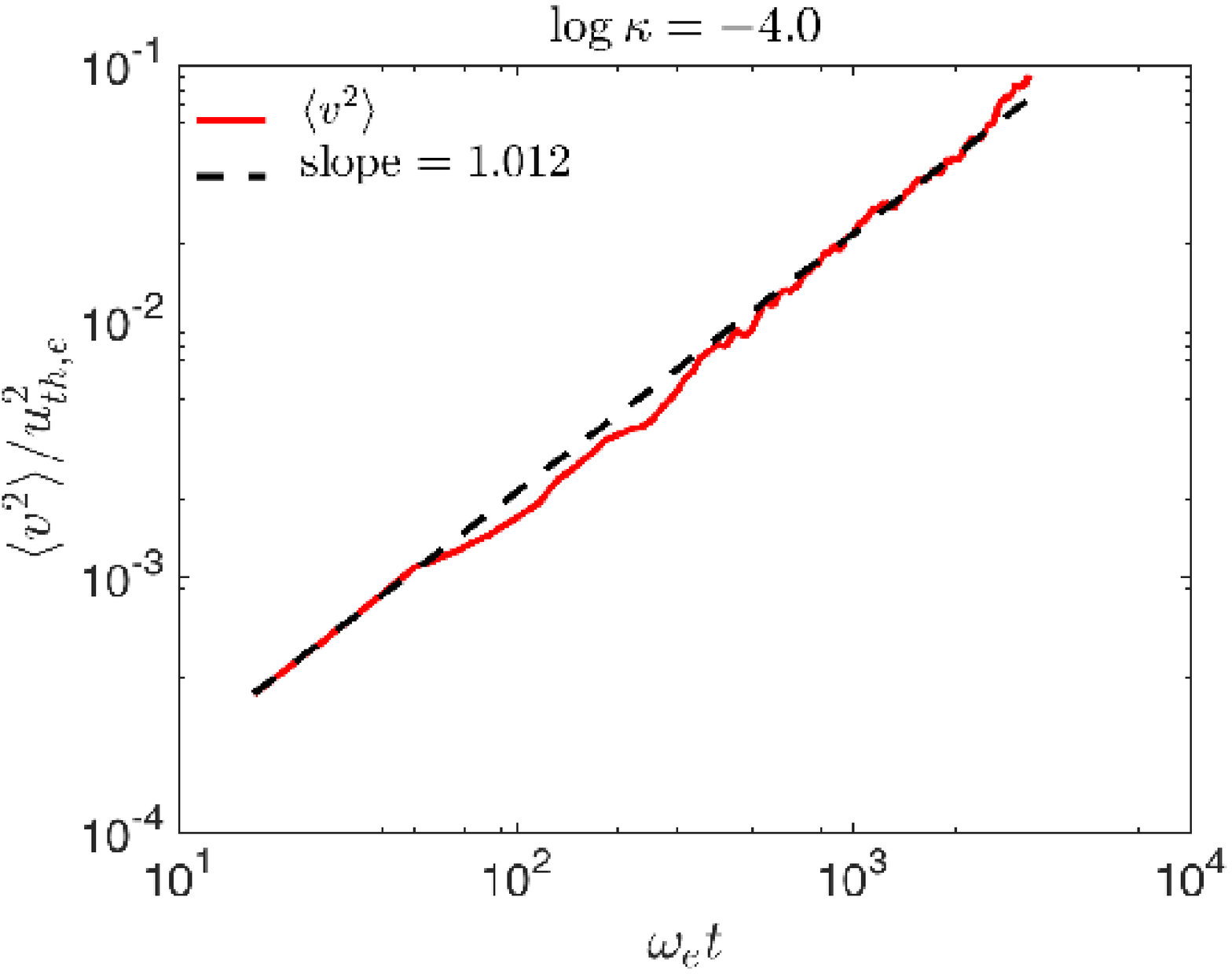}%
	\label{fig:vMSD -4}}\hfill
    \sidesubfloat[]{\includegraphics[width=0.4\columnwidth]{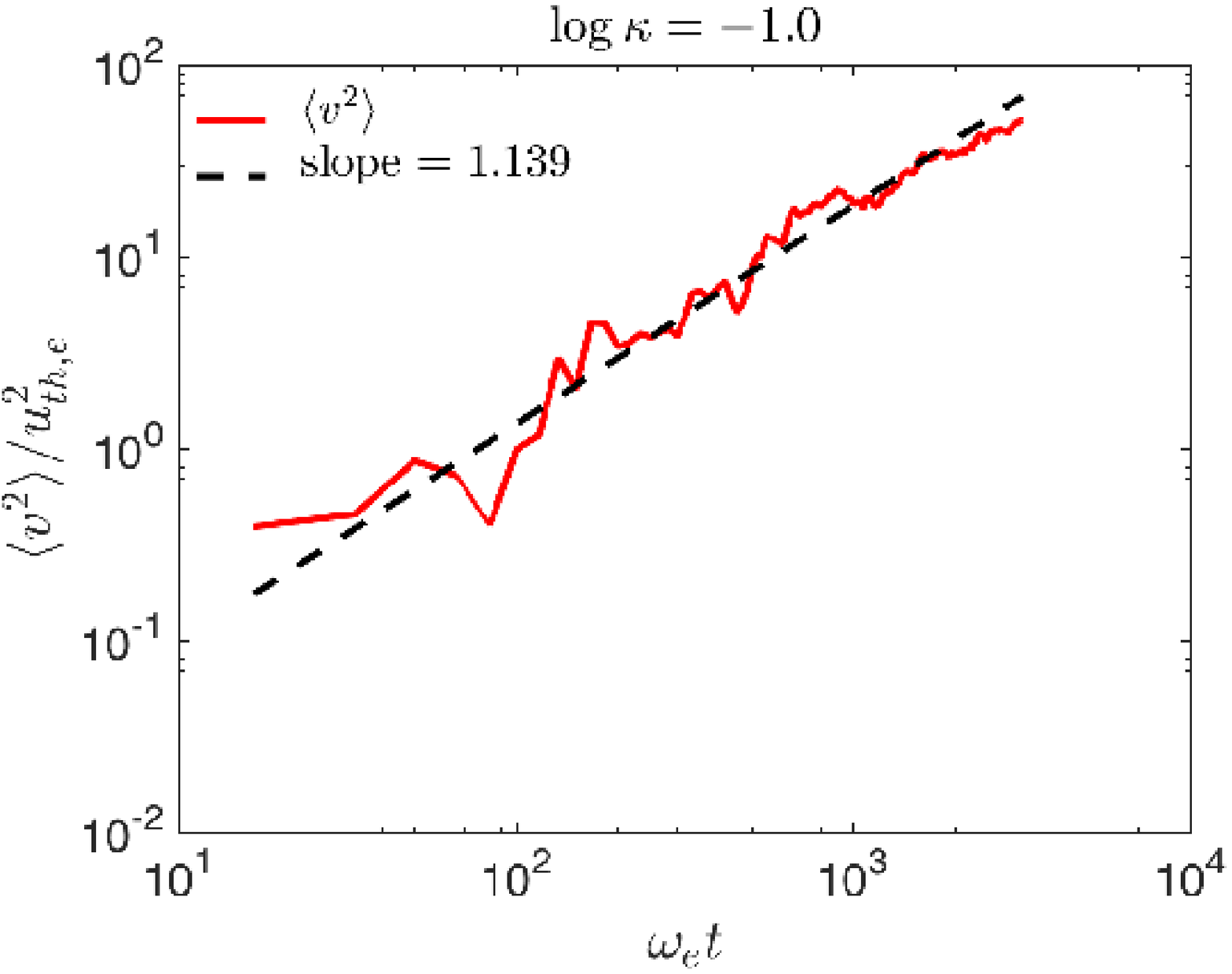}%
	\label{fig:vMSD -1}}\hfill
	\sidesubfloat[]{\includegraphics[width=0.4\columnwidth]{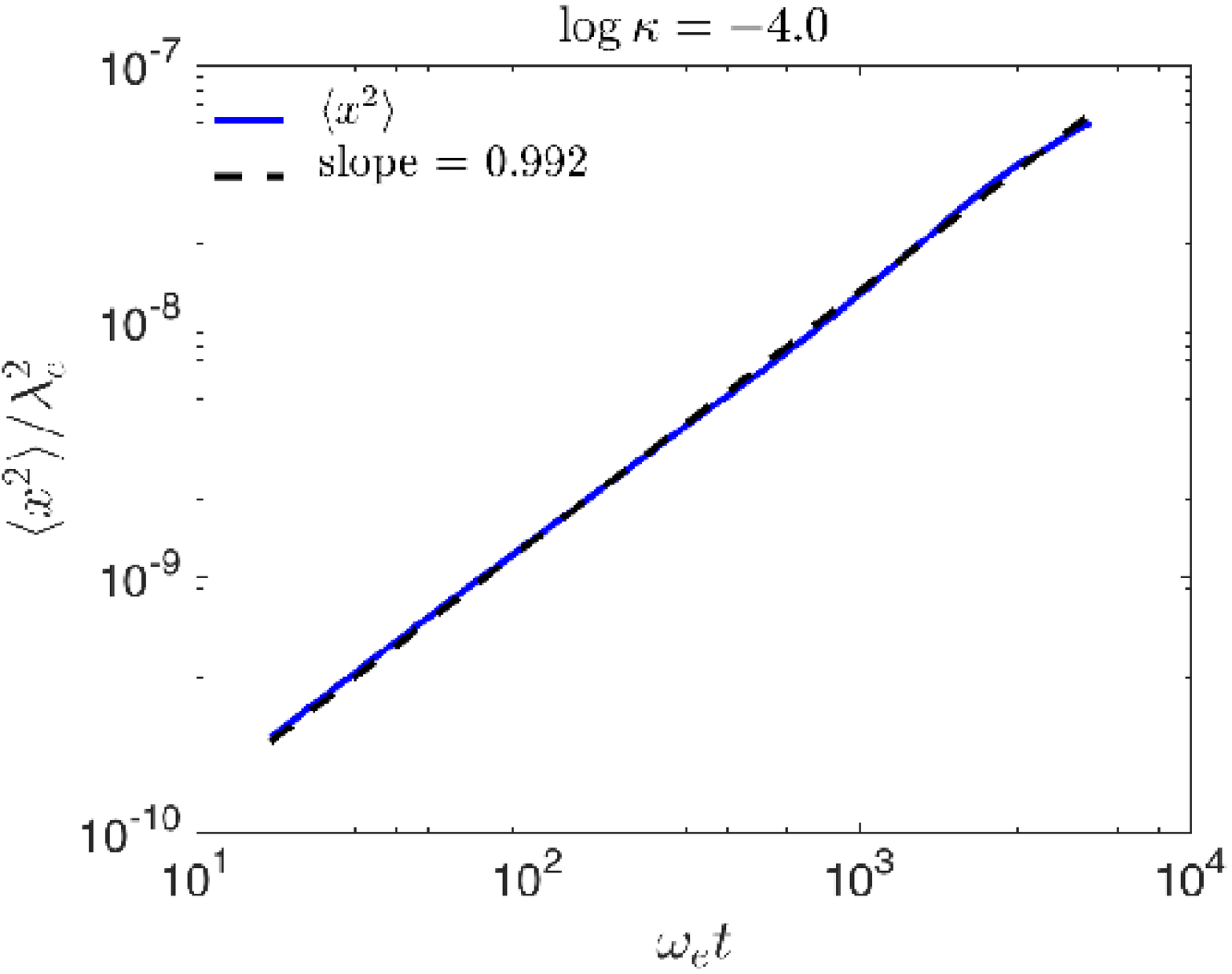}%
	\label{fig:xMSD -4}}\hfill
    \sidesubfloat[]{\includegraphics[width=0.4\columnwidth]{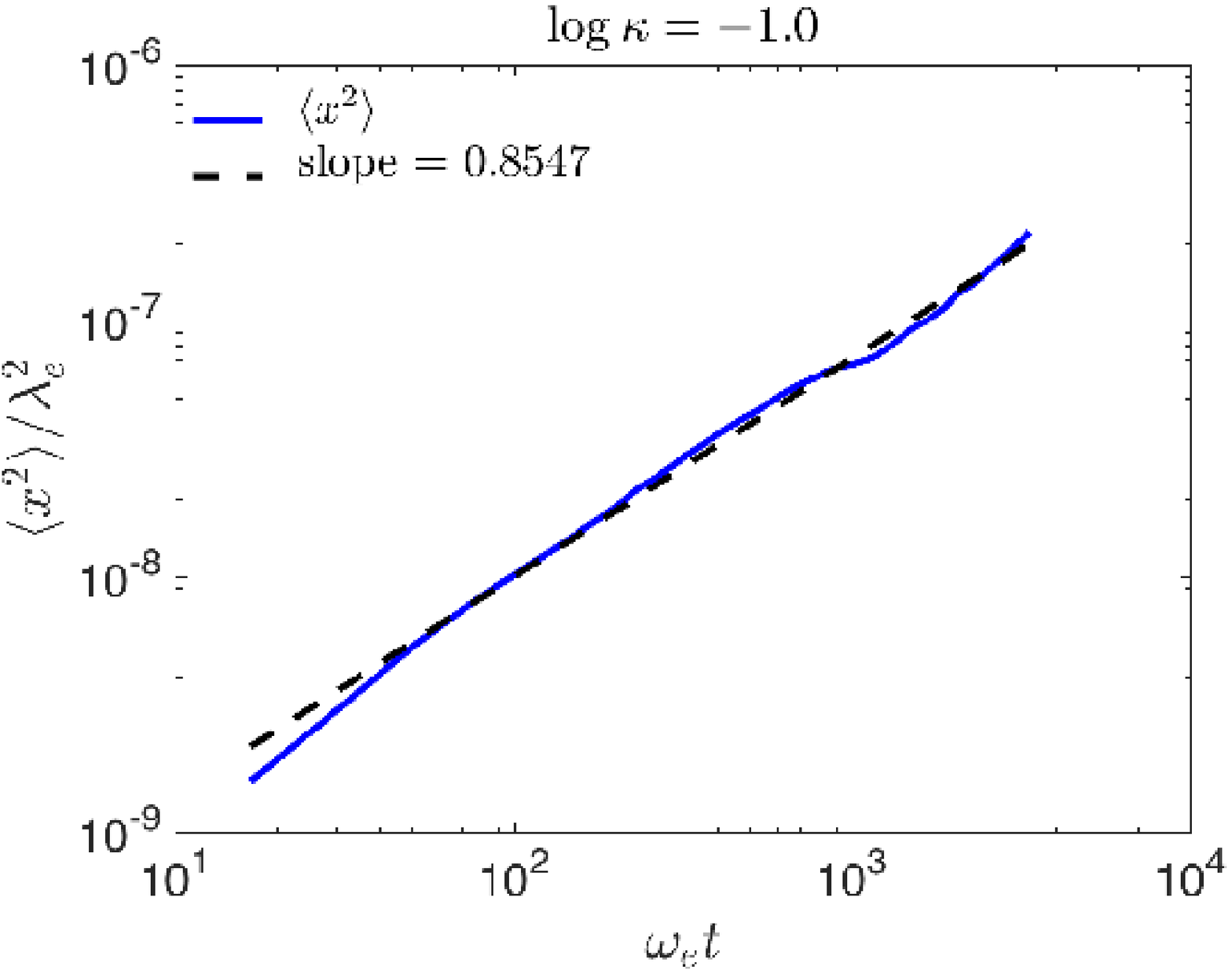}%
	\label{fig:xMSD -1}}\hfill
	\caption{The evolution over time of the energy of six randomly selected particles, normalized by their initial energy 
	(\textit{\protect\subref{fig:langmuir energies -4}}
	and \textit{\protect\subref{fig:langmuir energies -1}}), 
	the mean square displacement in velocity space
	(\textit{\protect\subref{fig:vMSD -4}}
	and \textit{\protect\subref{fig:vMSD -1}}), 
	and the mean square displacement in position space
   (\textit{\protect\subref{fig:xMSD -4}}
	and \textit{\protect\subref{fig:xMSD -1}}), 
    for $\kappa = 10^{-4}$, where QLT is expected to work (left column), and 
	for $\kappa = 10^{-1}$, where QLT is expected to be invalid (right column).
	}
	\label{fig:langmuir energies}
\end{figure}

  FIG.~\ref{fig:langmuir energies} shows the evolution of the energy of six, out of in total $N_p$, randomly selected particles. For the case in FIG.~\ref{fig:langmuir energies -4} that the waves carry energy equal to $0.01$\% of the total plasma thermal energy, the energy of the particles evolves in a stochastic, random walk like manner. If the wave energy increases to $10$\% of the total plasma thermal energy, as in FIG.~\ref{fig:langmuir energies -1}, 
  the evolution still is random walk like,
  the particles though experience abrupt energy jumps over short times, in some cases $2-3$ orders of magnitude larger than their initial energy, 
  Also note the much more extended dynamical range in FIG.~\ref{fig:langmuir energies -1} compared to
  FIG.~\ref{fig:langmuir energies -4}. 
  The random walk character of the evolution is visible on time-scales large enough so that the particles have entered the diffusive regime,
  and we find that
  the time needed for the particles to travel several tens of a typical wavelength is several hundreds of the plasma period.
  The diffusive time scale is also illustrated by the mean square displacements (MSD) in velocity and in position space, as 
  also shown in 
  FIG.~\ref{fig:langmuir energies}. In the case of low energy waves, the MSD in velocity and position space is proportional to time, which implies that the diffusion is normal in both spaces and the random walk is of classical nature (FIGS.\ \ref{fig:vMSD -4} and \ref{fig:xMSD -4}), whereas for the waves with larger energy content the diffusion process has become anomalous, namely super-diffusive in velocity space (FIG.\ \ref{fig:vMSD -1}) and sub-diffusive in position space (FIG.\ \ref{fig:xMSD -1}), which can clearly be attributed to nonlinear effects.

\begin{figure}[ht]
	\sidesubfloat[]{\includegraphics[width=0.45\columnwidth]{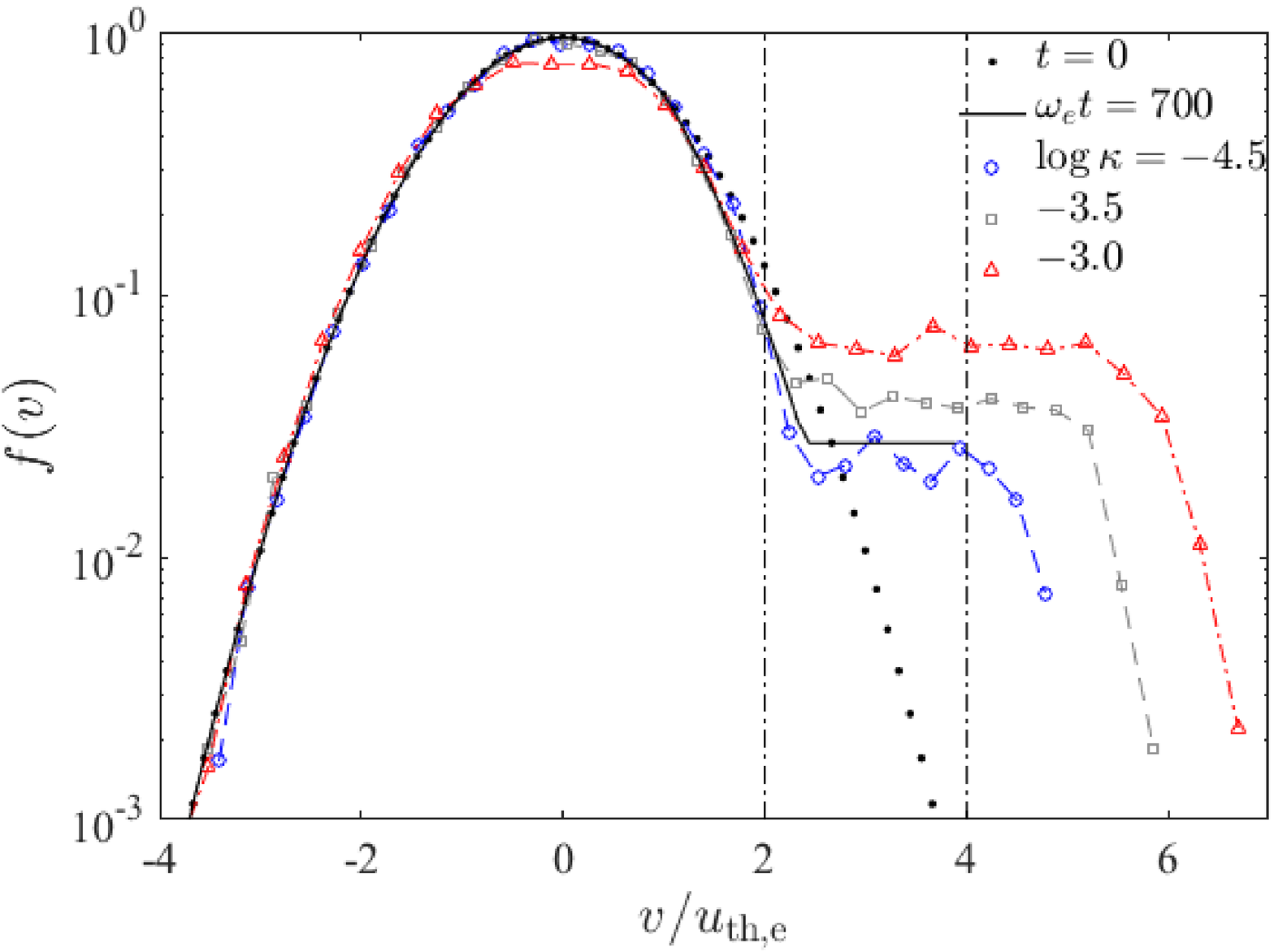}%
		\label{fig:first langmuir distributions}}
	\sidesubfloat[]{\includegraphics[width=0.45\columnwidth]{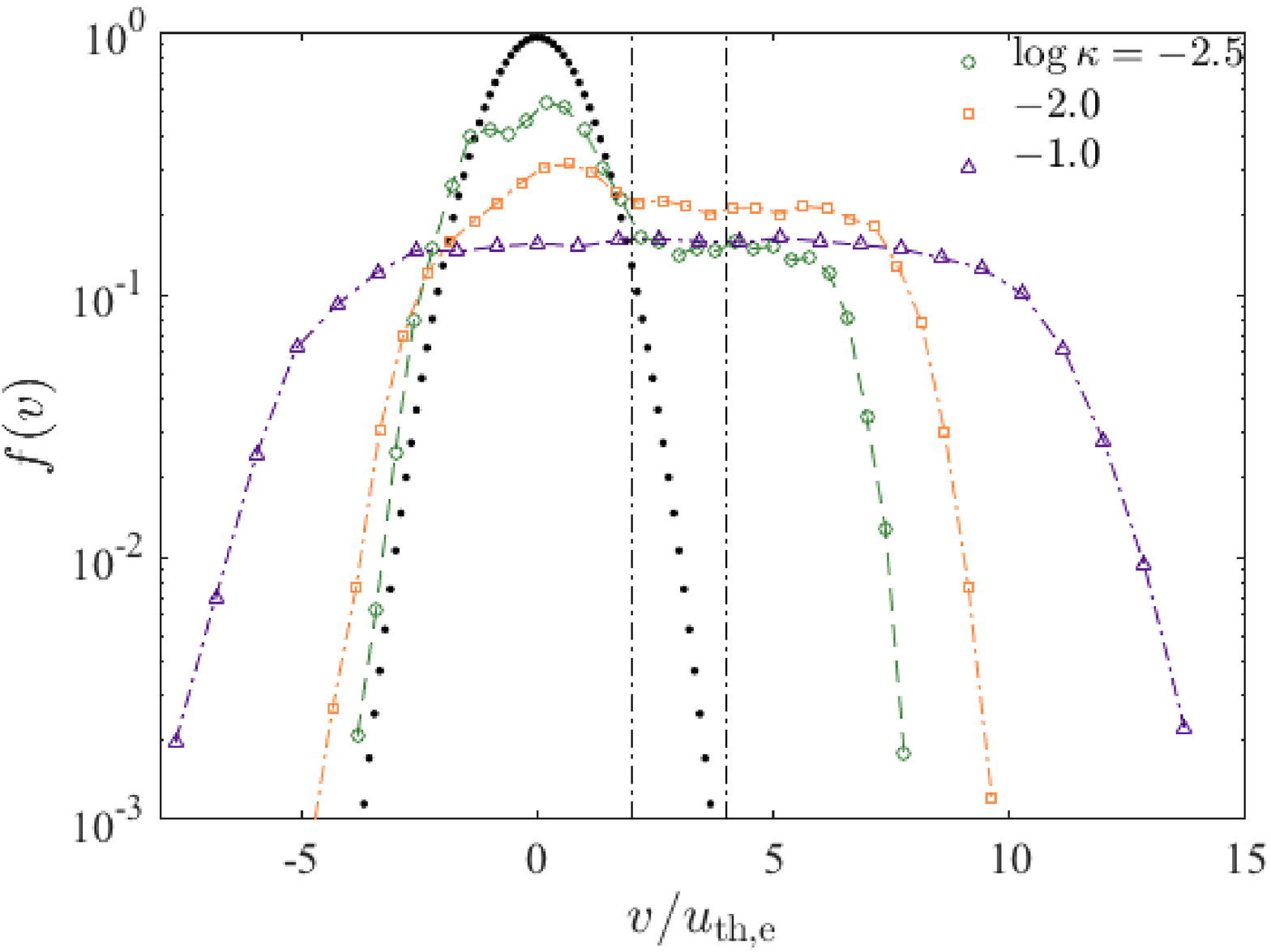}%
		\label{fig:second langmuir distributions}}
	\caption{\textit{\protect\subref{fig:langmuir diffusion coefficients}} For values of $\kappa < 10^{-3.5}$, the numerical results are consistent with QLT, since the plateau (black solid line) is predicted by Eq.~(\ref{eqthLM:plateau}) accurately. 
		\textit{\protect\subref{fig:langmuir current}} The initial distribution (black dotted line), for cases of $\kappa > 10^{-3.5}$, is modified beyond QLT's predictions. The vertical black dashed lines mark the minimum and maximum phase velocities of the excited waves.
	}
	\label{fig:langmuir distributions}
\end{figure}

The evolution of the particle velocity distribution function due to wave-particle interactions, for various values of $\kappa$, is shown in FIG.~\ref{fig:langmuir distributions}. According to Eq.~(\ref{eqthLM:quasi-stable condition}), we expect that the initial Maxwellian velocity distribution will evolve to form a plateau inside the resonance region, while practically no change will be observed in the non-resonant part of it, provided that the wave energy is restricted by $\log \kappa \ll -2.3 = \log \kappa _{\rm LM}^{\rm QL}$, according to condition (\ref{eqthLM:kappa restriction order}). As seen in FIG.~\ref{fig:first langmuir distributions}, for the range of wave energies with $\log \kappa < -3.5$ the velocity distribution's evolution is predicted accurately by QLT. Above this threshold the plateau is broadened and the non-resonant part of the distribution is also modified, as FIG.~\ref{fig:second langmuir distributions} demonstrates, in which cases of waves carrying energy up to $10$\% of the total plasma thermal energy are shown. The distribution in these cases is strongly modified and QLT fails to describe this modification. 

\begin{figure}[ht]
\sidesubfloat[]{\includegraphics[width=0.45\columnwidth]{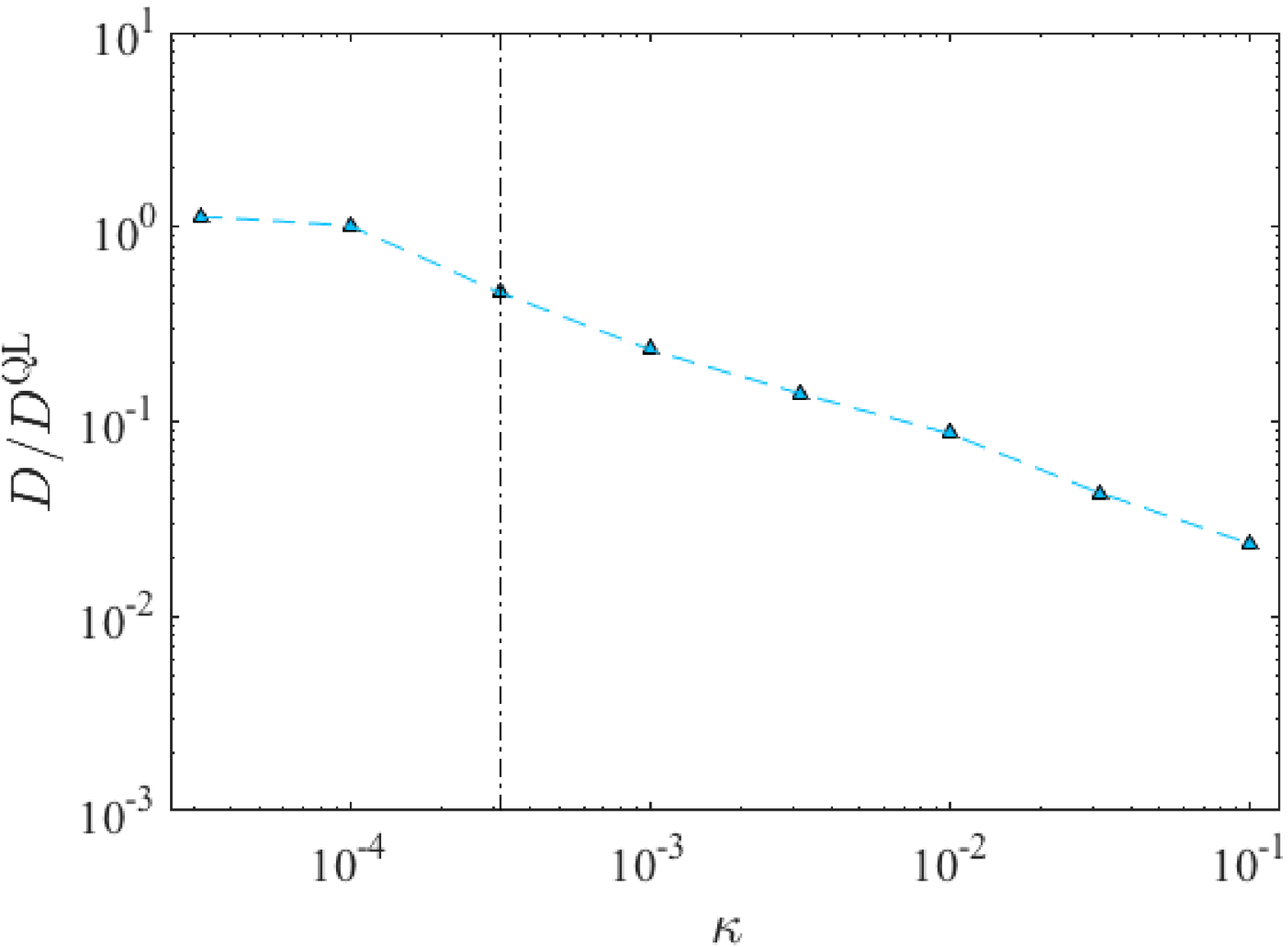}%
		\label{fig:langmuir diffusion coefficients}}
		\sidesubfloat[]{\includegraphics[width=0.45\columnwidth]{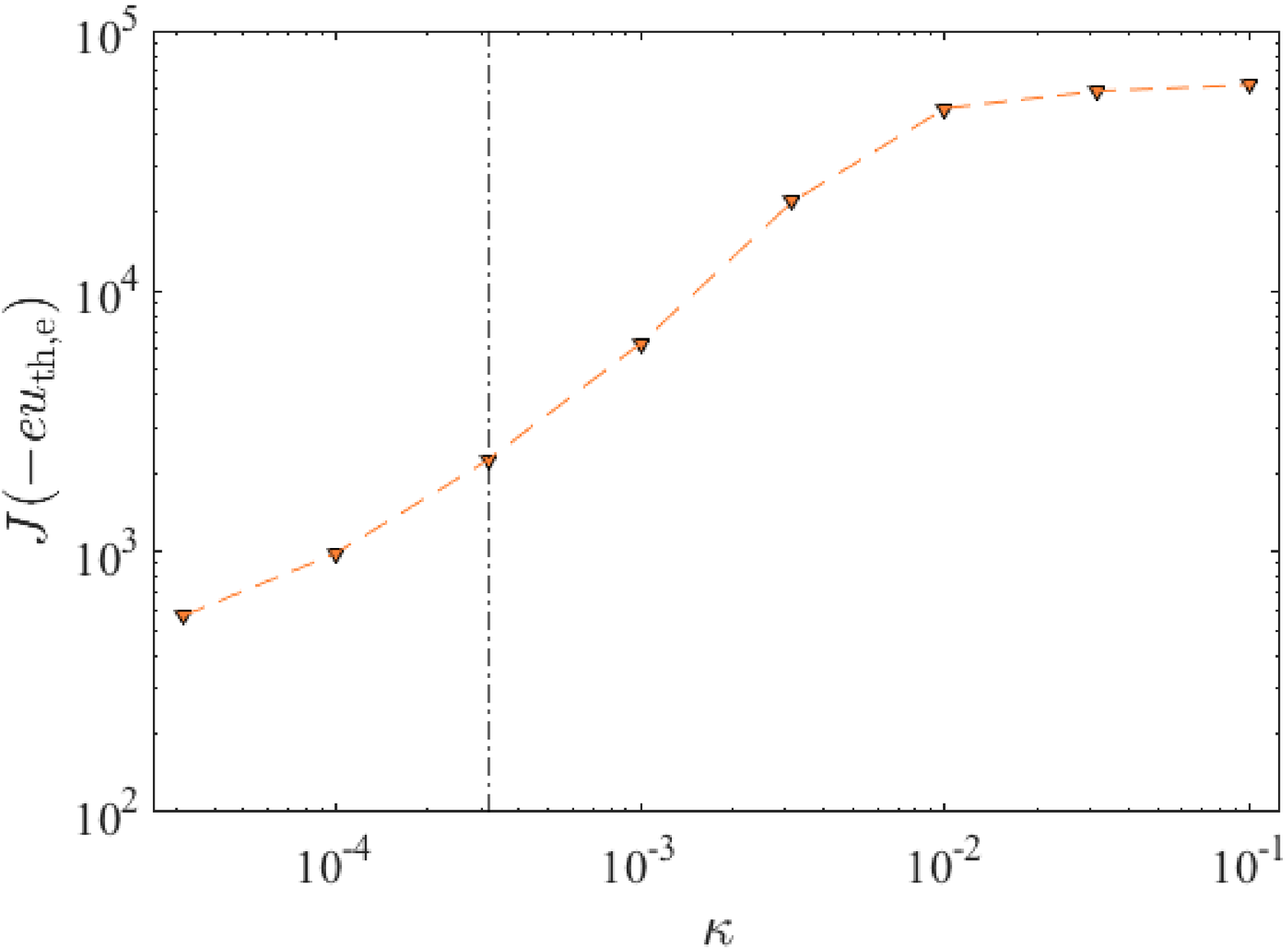}%
		\label{fig:langmuir current}}
	\caption{\textit{\protect\subref{fig:langmuir diffusion coefficients}}  Resonant test particle velocity diffusion coefficients according to Eq.\ (\ref{eqnum:diff coeff}), with $\delta t \simeq 30 \omega _e^{-1}$, normalized by the analytical expectation of the QLT. 
		   \textit{\protect\subref{fig:langmuir current}} Total electric current $J$ in the plasma, induced by the wave-particle interaction, in units of $-eu_{\rm th,e}$. 
		}
			
\end{figure}


FIG.~\ref{fig:langmuir diffusion coefficients} shows the comparison between the theoretical prediction of the velocity diffusion coefficients for resonant particles and the diffusivities that were obtained numerically. For the cases of waves with $\log \kappa < -3.5$ the diffusion coefficients are in good agreement, thus QLT provides accurate results. For $\log \kappa \gtrsim -3.5$ an increasing disagreement starts to appear, and the QLT overestimates the diffusion rates in the resonant velocity range. 
In these cases, the wave amplitudes are too high to satisfy QLT's 
prerequisites, the interaction is fully nonlinear and 
stochasticity comes into play in the non-resonant part of the phase space, so particles from the bulk of the velocity distribution diffuse very efficiently to very high velocities. This also leads to the flattening of the velocity distribution well outside the resonant part, which is very obvious in FIG.~\ref{fig:second langmuir distributions}. 

In FIG.~\ref{fig:langmuir current} the total electric current, $J$, as induced  by the wave-particle interaction, which breaks the isotropy of the initial Maxwellian, is shown as a function of $\kappa$. $J$ increases with increasing $\kappa$, until it starts to saturate above $\log \kappa \sim -2$. From then on, electrons with negative velocities are also noticeably accelerated, as can also be seen in Fig~\ref{fig:langmuir distributions}, hence the saturation of $J$.

Overall, we find that if $\log \kappa \lesssim -4$, QLT can be safely used as a valid description. The breaking point of the theory appears to be in the range of $\log \kappa \in [-4,-3.5]$, thus it is clearly $1.5$ orders of magnitude lower than the theoretical limit $\kappa _{\rm LM}^{\rm QL}$ in Eq.~(\ref{eqthLM:kappa restriction order}).

\section{Upper Hybrid and Lower Hybrid waves}
\label{sec5:hybrid}

Now we study the quasilinear theory's applicability for the case of purely perpendicularly propagating UH and LH waves. In this case, the turbulence is also electrostatic, and the resonances occur only in the perpendicular plane. These resonances result in the heating of the particles in $v_\perp$, while in $v_\|$ no significant energy gain is observed.

\subsection{Analytic predictions}
\label{subsec:hybrid theory}

As shown in Appendix~\ref{sec:appendixA-dispersion relations}, by making some assumptions, one can express the dispersion relations as $\omega_\mathbf{k} \simeq |\Omega_s|$, and $\gamma_\mathbf{k} \simeq -(\pi/4)(\omega_s^4/|\Omega_s|^3)$, where $s = e/i$ for UH/LH waves. Specifically, we have assumed small enough wave-numbers, and a strong magnetic field. Using the simplified expressions, it is easy to derive some important analytical results about the wave-particle interactions.

\begin{table}[ht]
	\caption{\label{tbl:hybrid parameters}Simulation parameters for the case of Upper Hybrid and Lower Hybrid waves.}
	\begin{ruledtabular}
		\begin{tabular}{lclc}
			\multicolumn{2}{c}{\bf Upper Hybrid} & \multicolumn{2}{c}{\bf Lower Hybrid}\\
			{\bf Parameter} & {\bf Value} & {\bf Parameter} & {\bf Value}\\
			$N_p$\footnotemark[1] & $2\times 10^4$ & $N_p$ & $2\times 10^4$\\
			$N_k$\footnotemark[1] & $100$ & $N_k$ & $100$\\
			$n_0({\rm cm}^{-3})$ & $10^9$ & $n_0({\rm cm}^{-3})$ & $10^9$\\
			$T({\rm eV})$ & $100$ & $T({\rm eV})$ $^{\rm a}$ & $100$\\
			$u_{\rm th,e}(c)$ & $1.4\times 10^{-2}$ & $u_{\rm th,i}(c)$ & $3.3\times 10^{-4}$\\
			$u_{\min}(u_{\rm th,e})$\footnotemark[2] & $3.4$ & $u_{\min}(u_{\rm th,i})$ & $2.5$\\
			$u_{\max}(u_{\rm th,e})$\footnotemark[2] & $4.0$ & $u_{\max}(u_{\rm th,i})$ & $3.0$\\
			$B_0({\rm G})$ & $500$ & $B_0({\rm kG})$ & $7$\\
		\end{tabular}
	\end{ruledtabular}
	\footnotetext[1]{Total number of particles and total number of exited waves}
	\footnotetext[2]{Minimum and maximum phase velocities of the wave spectrum}
\end{table}

More specifically, since the turbulence is electrostatic, with $\mathbf{E}_\mathbf{k}=E_\mathbf{k} \mathbf{k}/k_\perp \equiv E_\mathbf{k} \mathbf{\hat{e}}_\mathbf{x}$, from Eq.~(\ref{eqth:convenience function Epsilon}) one concludes that $\mathcal{E}^s_{n,\mathbf{k}}=nE_\mathbf{k} J_n(\zeta^s_\mathbf{k})/\zeta^s_\mathbf{k}$, where $\zeta^s_\mathbf{k} \equiv k_\perp v_\perp /\Omega_s$, and hence from Eq.~(\ref{eqth:velocity diff vector}) it follows that the vector
\begin{equation}\label{eqhybrid:diff vector}
\mathbf{d}^s_{n,\mathbf{k}}= \frac{nJ_n(\zeta^s_\mathbf{k})}{\zeta^s_\mathbf{k}} E_\mathbf{k} \mathbf{\hat{e}}_\perp .
\end{equation}
Thus, Eq.~(\ref{eqth:velocity diff tensor}) for the diffusion coefficients becomes
\begin{equation}\label{eqthUH:diff coeff}
D^{\rm QL}_{\perp s} \simeq \frac{\pi \omega_s^2}{2n_0m_s|\Omega_s|}\int d^3\mathbf{k} \; \mathcal{W}_\mathbf{k},
\end{equation}
as shown in Appendix~\ref{sec:appendixB-diffusion coefficients}, and which is just the purely perpendicular component and the only non-zero part of Eq.~(\ref{eqth:velocity diff tensor}). Resonances are purely in the perpendicular plane, and particles will be in gyro-resonance as long as the resonance condition $\omega_\mathbf{k}=n\Omega_s$ is satisfied, where $n$ is an integer. Also, we make sure that we select a strong enough magnetic field, such that the only resonance we observe is the $|n|=1$ in both cases.

Once again, the validity of quasilinear theory requires that the damping timescale $|\gamma_\mathbf{k}|^{-1}$ is much shorter than the averaged distribution's relaxation time $\tau_R$, 
as expressed by the condition in Eq.~(\ref{eqth:slow damping condition}), which here takes the form
\begin{equation}\label{eqthUH:kappa limit}
\kappa \ll \kappa_s^{\rm QL} \equiv \frac{n_0m_s\omega_s^2 \left[ \Delta (|\Omega_s|/k_\perp ) \right] ^2}{2 \Omega_s^2 W_{\rm tot}} ,
\end{equation}
where $s=e$ corresponds to the UH case and we denote the upper limit for QLT's validity according to the theory with $\kappa _{\rm UH}^{\rm QL}$, while the corresponding notation for the case of LH waves, with $s=i$, is $\kappa _{\rm LH}^{\rm QL}$. If this limit is satisfied, quasilinear theory is expected to be applicable, and the distribution function is expected to show heating. Also, in  case of applicability, quasilinear theory can make an estimate of the temperature. Specifically, by applying the transformation $\partial t'/\partial t=2D^{\rm QL}_{\perp s}$, we find as solution a Maxwellian with temperature 
\begin{equation}\label{eqthUH:perpendicular temperature increase}
\Delta T_s^{\rm QL} \simeq \frac{\pi \omega_s^2 W_{\rm tot}}{n_0|\Omega_s|} \kappa _s^{\rm QL} t,
\end{equation}
and QLT is valid if $\Delta T_s^{\rm QL}/m_s \ll u_{\rm th,s}^2$. 

\subsection{Numerical results}
\label{subsec:hybrid numerical results}

The parameters used in the simulations of the LH and UH wave cases are summarized in TABLE~\ref{tbl:hybrid parameters}. The kinetic energies of individual test particles and the average energy behave in a similar fashion as in the LW case (see FIG.~\ref{fig:langmuir energies}).

\begin{figure}[ht]
\sidesubfloat[]{\includegraphics[width=0.45\columnwidth]{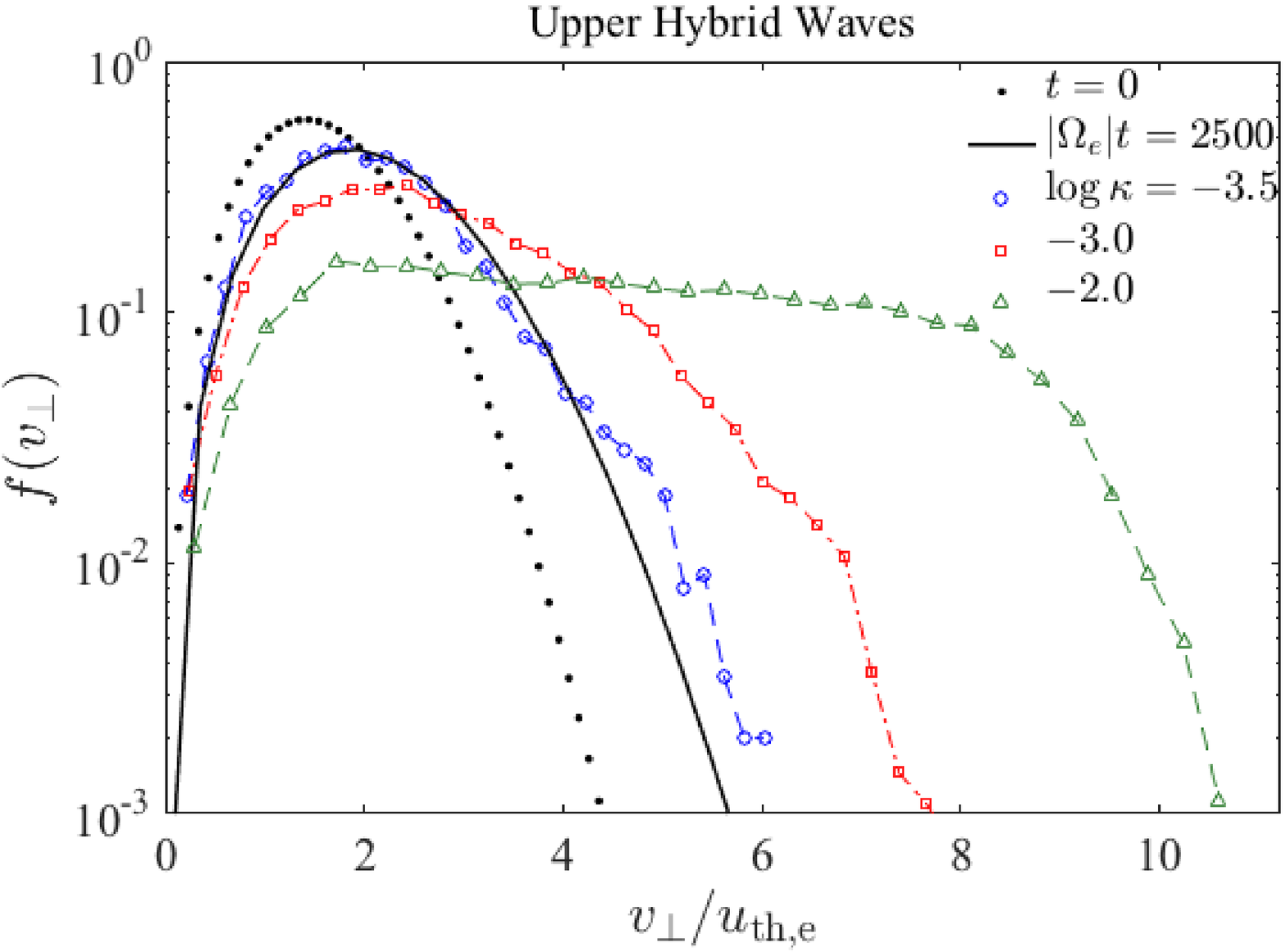}%
		\label{fig:upper hybrid distributions}}
		\sidesubfloat[]{\includegraphics[width=0.45\columnwidth]{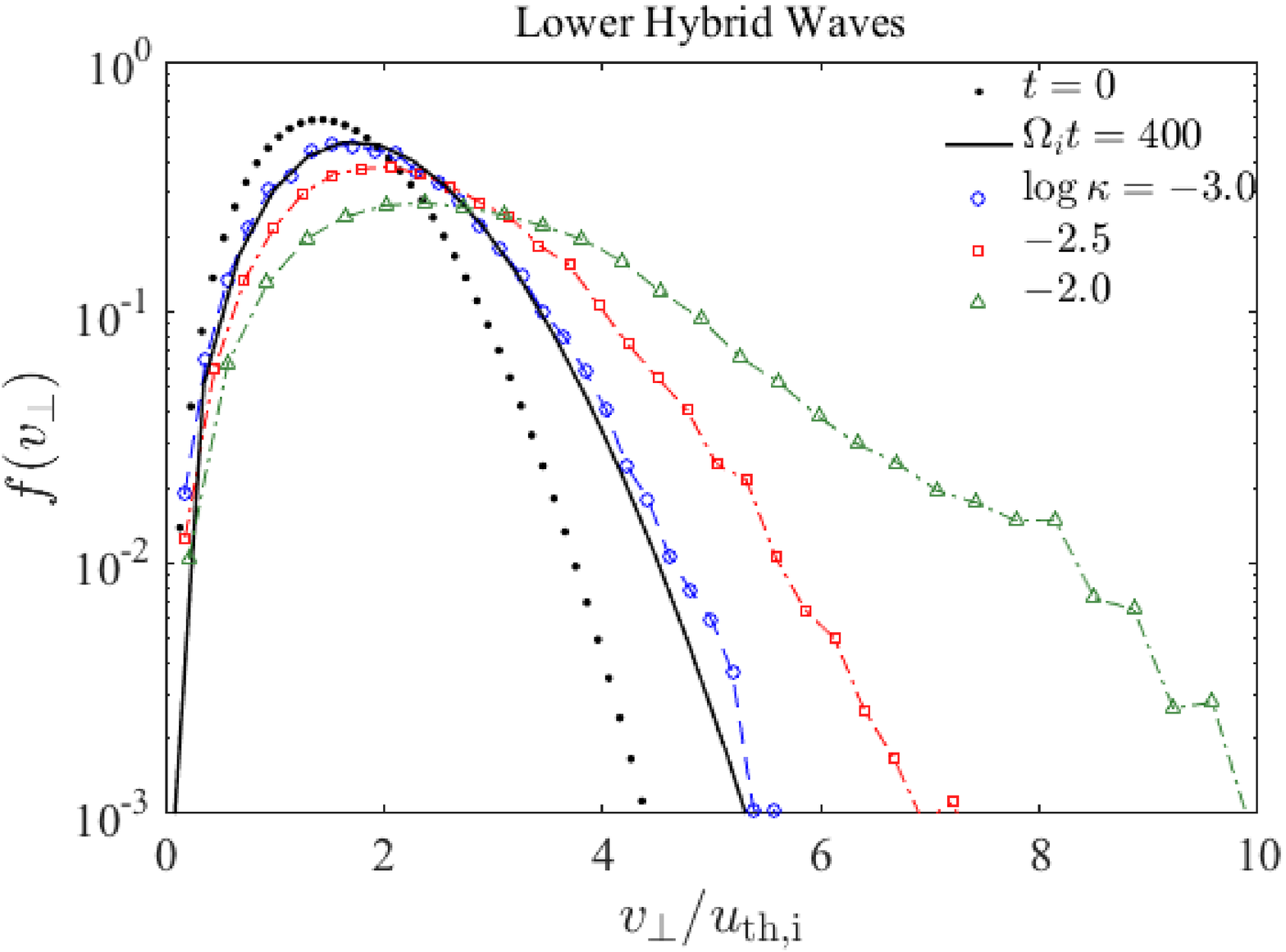}%
		\label{fig:lower hybrid distributions}}
	\caption{\textit{\protect\subref{fig:upper hybrid distributions}} Evolution of the perpendicular electron velocity distribution function in the interaction with an excited spectrum of Upper Hybrid waves, for various $\kappa$ values. 
	The final results correspond to $t=2500|\Omega_e|^{-1}$. \textit{\protect\subref{fig:lower hybrid distributions}} Evolution of the perpendicular ion velocity distribution function in the interaction with an excited spectrum of Lower Hybrid waves, for various $\kappa$ values. The final results correspond to $t=400\Omega_i^{-1}$.}
\end{figure}



The analytical expectation for the upper limit of wave energy for QLT's applicability in relation (\ref{eqthUH:kappa limit}) suggests that it must hold that $\log \kappa \ll -2.8 = \log \kappa _{\rm UH}^{\rm QL}$ for the case of UH waves, and $\log \kappa \ll -1.3 = \log \kappa _{\rm LH}^{\rm QL}$ for the case of LH waves, by using the parameters in TABLE~\ref{tbl:hybrid parameters}. If $\kappa$ is in the range suggested by these conditions, the velocity distribution function of the test particles is expected to show heating 
in the way predicted by Eq.~(\ref{eqthUH:perpendicular temperature increase}), and the diffusion coefficients can be approximated by Eq.~(\ref{eqthUH:diff coeff}).

\begin{figure}[ht]
\sidesubfloat[]{\includegraphics[width=0.45\columnwidth]{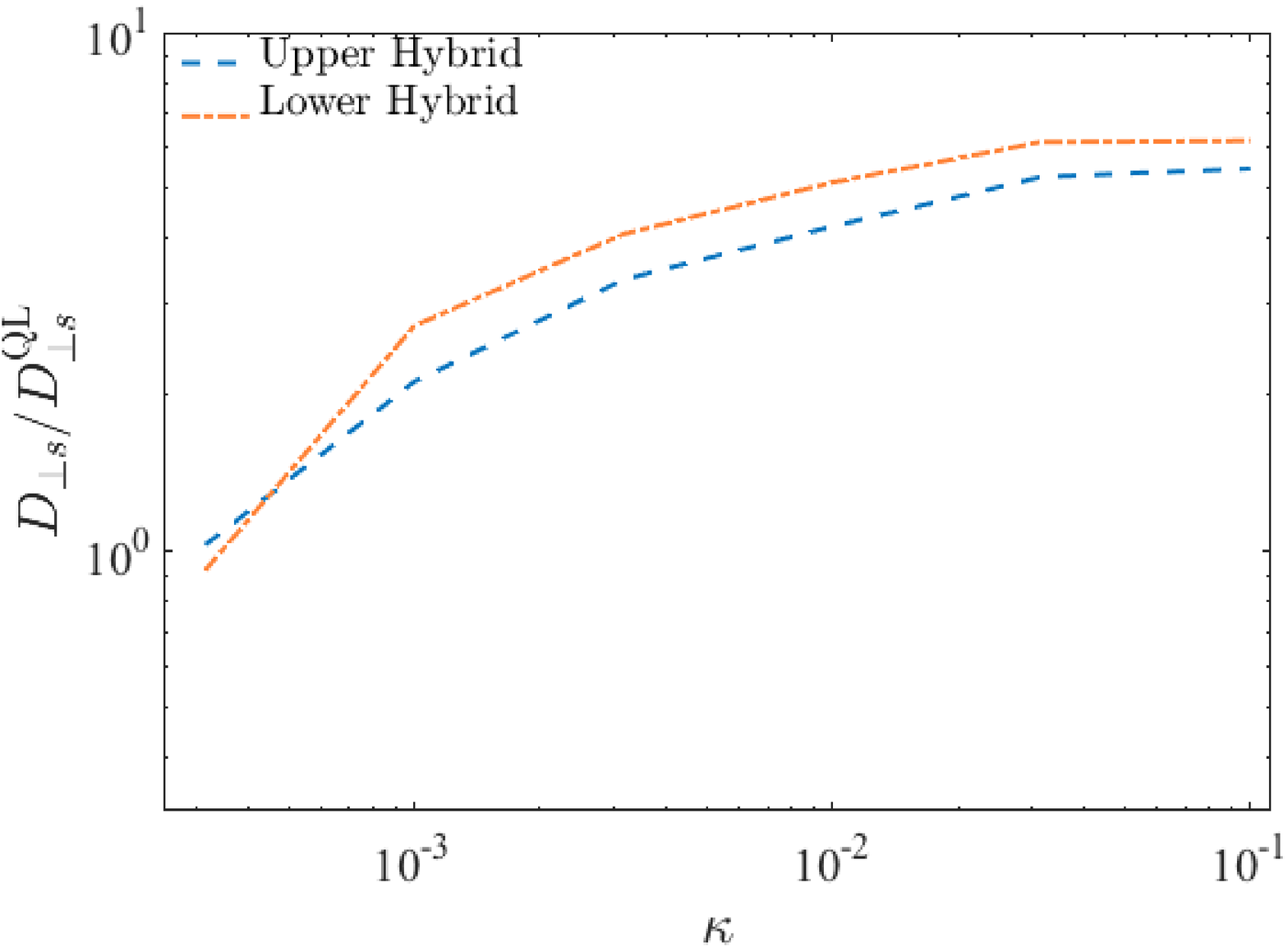}%
		\label{fig:hybrid diffusion coefficients}}
		\sidesubfloat[]{\includegraphics[width=0.45\columnwidth]{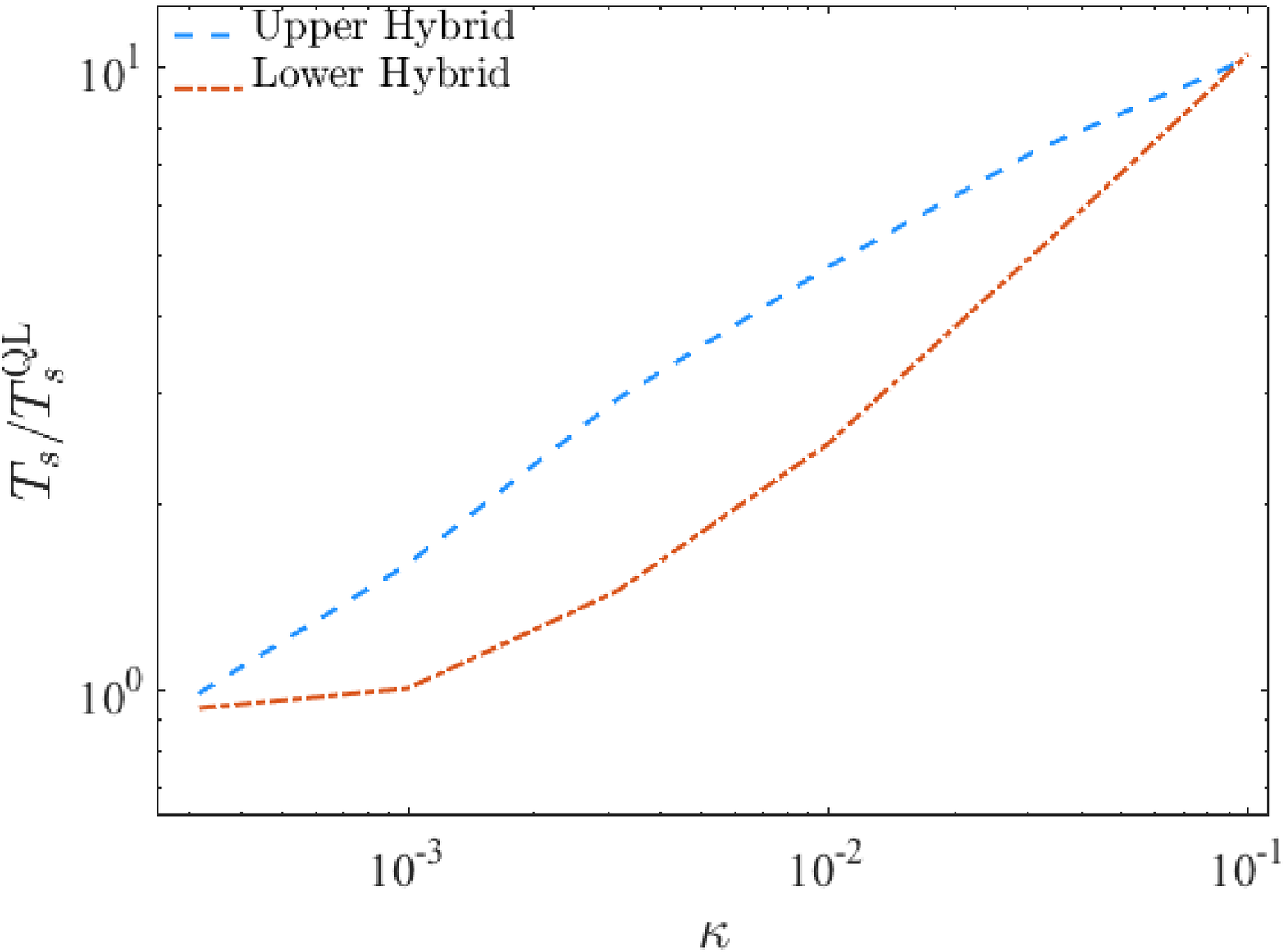}%
		\label{fig:hybrid temperatures}}
	\caption{\textit{\protect\subref{fig:hybrid diffusion coefficients}}  Test electron and ion velocity diffusion coefficients, using the definition in Eq.\ (\ref{eqnum:diff coeff}), with $\delta t \simeq 200 |\Omega_e|^{-1}$ and $\delta t \simeq 32\Omega_i^{-1}$ for the UH and LH waves, respectively, normalized by the analytically expected value, and for varying $\kappa$. 
	\textit{\protect\subref{fig:hybrid temperatures}} Comparison between the temperature of the final distribution as predicted by Eq.~(\ref{eqthUH:kappa limit}) and the numerical results, for the cases of Upper Hybrid and Lower Hybrid waves, and for varying $\kappa$. }
			
\end{figure}


In FIG.~\ref{fig:upper hybrid distributions}, the evolution of the perpendicular electron velocity distribution function $f(v_\perp)$  is shown for the case of UH waves. The heating of test-particles 
for $\log \kappa = -3.5$ is consistent with QLT's prediction (Eq.~(\ref{eqthUH:perpendicular temperature increase})). If the wave energy is increased above this value, QLT underestimates the heating, and for $\log \kappa \geq -3$, the theory is not valid. 
Also, the results of the comparison between the numerical and analytical diffusion coefficients, shown in FIG.~\ref{fig:hybrid diffusion coefficients}, confirm this limit,
as $\kappa$ increases, an increasing disagreement between the numerical and theoretical results appears, and QLT underestimates the diffusion coefficients for $\kappa \gtrsim 10^{-3}$.
Furthermore, in FIG.~\ref{fig:hybrid temperatures} the temperatures of the final distributions 
compared to Eq.~(\ref{eqthUH:perpendicular temperature increase}) are shown. As can be seen, the results for the temperature indicate the same limit for $\kappa$,
QLT underestimates the final temperature 
for $\kappa \gtrsim 10^{-3}$.
Overall, the results suggest the wave energy range of $\log \kappa \lesssim -3.5$ for QLT's applicability in the UH case, and that in the interval $\log \kappa \in [-3.5,-3]$ the first signs of QLT's invalidity can be found. Thus, 
the maximum value of $\kappa$ for QLT's validity is less than an order of magnitude lower than the theoretical $\kappa _{\rm UH}^{\rm QL}$.

The final perpendicular velocity distributions of the ions, after their interaction with an excited spectrum of LH waves, for various $\kappa$ values, is shown in FIG.~\ref{fig:lower hybrid distributions}. In this case, the theoretical heating
(Eq.~(\ref{eqthUH:perpendicular temperature increase}))
is consistent with the numerical results for $\log \kappa \lesssim -3$, and fails to describe the numerical results for larger values of the wave energy, by underestimating the final temperature. 
This also is obvious from FIG.~\ref{fig:hybrid diffusion coefficients}, in which one can see that the analytical diffusion coefficients are systematically lower than the corresponding numerical ones if $\log \kappa > -3$. 
That there is also underestimation of the heating by QLT 
can be seen in FIG.~\ref{fig:hybrid temperatures}, from which the same limit for $\kappa$ can be inferred. 
Thus, in the LH case, the upper limit for QLT's validity can be estimated to lie in $\log \kappa \in [-3,-2.5]$, hence it is more than an order of 
magnitude lower than 
the theoretical $\kappa _{\rm LH}^{\rm QL}$.

\section{Summary}
\label{sec:6-Conclusions}

In this article, we explored the limitations of the QLT for the interaction of electrostatic waves (LW, UH, LH) with the plasma. We used a spectrum of waves with energy $W_0=\kappa W_{tot}$, where $W_{tot}$ is the total thermal energy of the plasma, and a test particle numerical code to analyze and search for the transition from the QLT to the non-linear evolution of the test-particles.

Our main results are the following:

\begin{enumerate}
	\item For the LW case, using the basic criterion for the validity of the quasilinear approximation, i.e.\ the relaxation time of the particle evolution should be much shorter than the damping time of the waves, we estimated the maximum wave energy ($\kappa_{QL}$) below which the QLT is valid. 
	\item We estimated the diffusion coefficients analytically and numerically and demonstrated that when $\kappa>\kappa_{QL}$ then $D>D_{QL}.$ Also, the current drive of the electrons induced by the waves increases drastically in the range where the QLT breaks down.
	\item We repeated the above analysis for $UH$ and $LH$ waves, and we determined the limit
	$\kappa_{QL}$ below which the numerical and analytical results agree. We estimated diffusion coefficients and the rate of heating of the electrons and ions in the presence of low and strong amplitude UH and LH waves, respectively.
	
\end{enumerate}

It would be useful to repeat our calculation with the use of PIC simulations and for applications more closely related with specific laboratory and space plasma settings. 


\begin{acknowledgments}
We thank the anonymous referee for helpful comments.
This work was supported by (a) the national Programme for the Controled Thermonuclear Fussion, Hellenic Republic, (b) The EJP Cofund Action SEP-210130335 EUROfusion. The sponsors do not bear any responsibility for the content of this work.
\end{acknowledgments}

\begin{appendix}

\section{Dispersion relations for Upper Hybrid and Lower Hybrid waves}
\label{sec:appendixA-dispersion relations}

In order to calculate the dispersion relations of UH and LH waves, we insert the solutions for $\delta f^s_\mathbf{k}$, expressed as in (\ref{eqth:deltaf analytic solution}), into Poisson's equation (\ref{eqth:Poisson equation linear}), assuming that the zeroth order density of ions and electrons are equal, $n_{0,i}=n_{0,e}$, and we consider purely perpendicular propagation of $\delta \mathbf{E}_\mathbf{k}$, so that $\mathbf{k} = k_\perp \hat{\mathbf{e}}_\perp$, which gives
\begin{widetext}
	\begin{eqnarray} \label{eqthapp:Poisson equation}
	ik_\perp \delta E_\mathbf{k} = -i \sum\limits_{s=i,e} \frac{4\pi e^2}{m_s}  \sum \limits _{n,m=-\infty}^\infty \delta E_\mathbf{k} \int _0^\infty v_\perp dv_\perp \left( \frac{n}{\omega _\mathbf{k}-n\Omega_s} \right) \frac{J_m(\zeta^s_\mathbf{k})J_n(\zeta^s_\mathbf{k})}{\zeta^s_\mathbf{k}}  &&\\ \int _0^{2\pi}d\phi \; e^{i(m-n)\phi} \; \frac{\partial}{\partial v_\perp}\int _{-\infty}^\infty dv_\| \; f^s_0(\mathbf{v}),	\nonumber 
	\end{eqnarray}
\end{widetext}
where $\zeta ^s_\mathbf{k}\equiv k_\perp v_\perp /\Omega_s$.
	
We then insert the Maxwellian distribution (\ref{eqintr:Initial Maxwellian}) into Eq.~(\ref{eqthapp:Poisson equation}), and using the integral identities \cite{abra72}
\begin{equation}
\int _0^{2\pi}e^{i(m-n)\phi}d\phi =2\pi \delta _{nm} \; , \; \; \int _{-\infty}^\infty e^{-x^2}dx=\sqrt{\pi} \; , \nonumber
\end{equation}
\begin{eqnarray}
\int _0^\infty xe^{-p^2x^2}&J&_n(ax)J_n(bx) dx \nonumber \\
&& = \dfrac{1}{2p^2}\exp \left( -\frac{a^2+b^2}{4p^2} \right) I_n\left( \frac{ab}{2p^2} \right) , \nonumber
\end{eqnarray}
we end up with the relation
\begin{eqnarray}\label{eqapp:first form of dispersion relation UH}
1&-&\sum \limits_{s = i,e} \frac{\omega_s^2}{\Omega_s}\frac{e^{-\xi^s_k}}{\xi^s_k} \sum \limits _{n=-\infty}^\infty nI_n(\xi^s_k) \nonumber \\
&&\times \left[ \mathcal{P}\frac{1}{\omega_\mathbf{k}-n\Omega_s}-i\frac{\pi}{|\Omega_s|}\delta \left( n-\frac{\omega_k}{\Omega_s} \right) \right] =0,
\end{eqnarray}
where $I_n(z)\equiv i^{-n}J_n(iz)$ are the modified Bessel functions of first kind, and $\xi^s_\mathbf{k}\equiv (k_\perp u_{\rm th,s}/\Omega_s)^2$.
The symbol $\mathcal{P}$ denotes the principal value.

The principal value in the summation in Eq.~(\ref{eqapp:first form of dispersion relation UH}) can be written as
\begin{widetext}
\begin{eqnarray}\label{eqapp:initial summation transform}
\sum \limits _{n=-\infty}^\infty \frac{nI_n(\xi^s_\mathbf{k})}{\omega_\mathbf{k} -n\Omega_s} = \sum \limits _{n=-\infty}^\infty \frac{n(\omega_\mathbf{k} +n\Omega_s)I_n(\xi^s_\mathbf{k})}{\omega_\mathbf{k}^2-n^2\Omega_s^2} &=& \frac{1}{2}\xi^s_\mathbf{k}\Omega_s\sum \limits _{n=-\infty}^\infty \left[ \frac{ (n-1)I_{n-1} }{\omega_\mathbf{k}^2-n^2\Omega_s^2} +\frac{ -(n+1)I_{n+1} }{\omega_\mathbf{k}^2-n^2\Omega_s^2}+\frac{ I_{n-1}+I_{n+1} }{\omega_\mathbf{k}^2-n^2\Omega_s^2} \right] \nonumber \\
& \equiv & \frac{1}{2}\xi^s_\mathbf{k}\Omega_s(\mathcal{S}_1+\mathcal{S}_2+\mathcal{S}_3)
\end{eqnarray}
\end{widetext}
after using the Bessel functions recurrence relation $I_n(z)=(z/2n)[I_{n-1}(z)-I_{n+1}(z)]$ and the property $I_{-n}(z)=I_n(z)$. For convenience, we analyze each summation in (\ref{eqapp:initial summation transform}) separately.

\noindent
$\blacktriangleright$ For the first one, we have
\begin{eqnarray}\label{eqapp:1st sum}
\mathcal{S}_1 &=& \sum \limits _{n=-\infty}^\infty \frac{ nI_n }{(\omega_\mathbf{k}^2-\Omega_s^2)-n(n+2)\Omega_s^2}\nonumber \\
&=& -\sum \limits _{n=-1,0}\frac{(n+2)I_{n+2}}{(\omega_\mathbf{k}^2-\Omega_s^2)-n(n+2)\Omega_s^2}\nonumber \\
&& +\sum \limits _{n=1}^\infty\frac{nI_n-(n+2)I_{n+2}}{(\omega_\mathbf{k}^2-\Omega_s^2)-n(n+2)\Omega_s^2}.
\end{eqnarray}

\noindent
$\blacktriangleright$ The second one becomes
\begin{eqnarray}\label{eqapp:2nd sum}
\mathcal{S}_2 &=& \sum \limits _{n=-\infty}^\infty \frac{ -(n+2)I_{n+2} }{(\omega_\mathbf{k}^2-\Omega_s^2)-n(n+2)\Omega_s^2} = -\frac{2I_2}{\omega_\mathbf{k}^2-\Omega_s^2}-\frac{I_1}{\omega_\mathbf{k}^2} \nonumber \\
&& +\sum\limits_{n=1}^\infty \frac{nI_n-(n+2)I_{n+2}}{(\omega_\mathbf{k}^2-\Omega_s^2)-n(n+2)\Omega_s^2}=\mathcal{S}_1.
\end{eqnarray}

\noindent
$\blacktriangleright$ For the third one, and with the aid of the relation
\begin{equation}
\sum \limits _{n=-\infty}^\infty \frac{ I_{n \mp 1} }{\omega_\mathbf{k}^2-n^2\Omega_s^2}=\sum \limits _{n=-\infty}^\infty \frac{ I_n }{(\omega_\mathbf{k}^2-\Omega_s^2)-n(n\pm 2)\Omega_s^2}, \nonumber
\end{equation}
we get
\begin{eqnarray}\label{eqapp:3rd sum}
\mathcal{S}_3 &=& \sum \limits _{n=-\infty}^\infty \frac{ I_n+I_{n+2} }{(\omega_\mathbf{k}^2-\Omega_s^2)-n(n+2)\Omega_s^2} \nonumber \\
&=&\frac{I_0+I_2}{\omega_\mathbf{k}^2-\Omega_s^2} +\sum \limits_{n=-1,0}\frac{I_n+I_{n+2}}{(\omega_\mathbf{k}^2-\Omega_s^2)-n(n+2)\Omega_s^2} \nonumber \\
&&+2\sum \limits_{n=1}^\infty \frac{I_n+I_{n+2}}{(\omega_\mathbf{k}^2-\Omega_s^2)-n(n+2)\Omega_s^2} \nonumber \\
&=& 2 \left[ \left( \frac{I_0+I_2}{\omega_\mathbf{k}^2-\Omega_s^2}+\frac{I_1}{\omega _\mathbf{k} ^2} \right) \right. \nonumber \\
&& \left. +\sum \limits_{n=1}^\infty \frac{I_n+I_{n+2}}{(\omega_\mathbf{k}^2-\Omega_s^2)-n(n+2)\Omega_s^2} \right] .
\end{eqnarray}
After plugging (\ref{eqapp:initial summation transform})-(\ref{eqapp:3rd sum}) into  (\ref{eqapp:first form of dispersion relation UH}), we end up with the relation 
\begin{widetext}
\begin{equation}\label{eqapp:general dispersion relation}
1 - \sum \limits_{s=i,e} \omega_s^2e^{-\xi^s_\mathbf{k}}\left\{ \frac{I_0-I_2}{\omega_\mathbf{k}^2-\Omega_s^2} + \left[ \sum \limits_{n=1}^\infty \frac{(n+1)(I_n-I_{n+2})}{\omega_\mathbf{k}^2-(n+1)^2\Omega_s^2} \right] \right\} = - i\pi \sum \limits_{s=i,e} \frac{\omega_s^2}{\Omega_s}\frac{e^{-\xi^s_\mathbf{k}}}{\xi^s_\mathbf{k}}\sum \limits _{n=-\infty}^\infty nI_n \delta \left( \omega_\mathbf{k}-n\Omega_s \right) .
\end{equation}
\end{widetext}

The above relation can be significantly simplified if we consider that $\omega_s/|\Omega_s|\ll 1$ and also that $\xi^s_\mathbf{k}=(k_\perp\lambda_s)^2(\omega_s/\Omega_s)^2\ll 1$, where $\lambda_s = u_{\rm th,s}/\omega_s$ is the Debye length. Then we can approximate \cite{abra72}
$$I_n(\xi^s_\mathbf{k})\sim \frac{(\xi^s_\mathbf{k}/2)^n}{\Gamma (n+1)} \; , \; \; \text{for} \; \; \xi^s_\mathbf{k} \ll 1,$$
and every $I_{n>0}$ is much smaller than $I_0$. Hence, the term in the summation on the l.h.s. of Eq.~(\ref{eqapp:general dispersion relation}) that is in square brackets vanishes when compared to the term to its left, and therefore can be neglected. Also, since the following relation holds true \cite{abra72}
$$1=e^{-\xi^s_\mathbf{k}}\sum \limits _{n=-\infty}^\infty I_n(\xi^s_\mathbf{k}) \simeq e^{-\xi^s_\mathbf{k}}I_0(\xi^s_\mathbf{k}) \; , \; \; \text{for} \; \; \xi^s_\mathbf{k}\ll 1,$$
we can make the approximation
\begin{equation}\label{eqapp:relation for real freq}
\frac{e^{-\xi^s_\mathbf{k}}(I_0-I_2)}{\omega_\mathbf{k}^2-\Omega_s^2} \simeq \frac{e^{-\xi^s_\mathbf{k}}I_0}{\omega_\mathbf{k}^2-\Omega_s^2} \simeq \frac{1}{\omega_\mathbf{k}^2-\Omega_s^2}.
\end{equation}

Finally, to find the solution for $\omega_\mathbf{k}$, we equate the real part of (\ref{eqapp:general dispersion relation}) to zero, after taking (\ref{eqapp:relation for real freq}) into account, which yields the dispersion relation
\begin{equation}\label{eqapp:real freq equation}
1-\frac{\omega_e^2}{\omega_\mathbf{k}^2-\Omega_e^2}-\frac{\omega_i^2}{\omega_\mathbf{k}^2-\Omega_i^2}=0.
\end{equation}

For the case of UH waves, in which the ion contribution is neglected, Eq.~(\ref{eqapp:real freq equation}) gives the solution $\omega_\mathbf{k}=|\Omega_e| [1+(\omega_e/\Omega_e)^2]^{1/2}\simeq |\Omega_e|$, under the condition $\omega_e\ll |\Omega_e|$, which we also assume. We then insert the imaginary part of $\sigma_\mathbf{k}$, namely $\gamma_\mathbf{k}$, and express the l.h.s. of (\ref{eqapp:general dispersion relation}) as
\begin{equation}\label{eqapp:relation for imaginary part}
1-\frac{\omega_e^2}{\omega _\mathbf{k}^2-\Omega_e^2} \simeq -i\frac{2\omega_\mathbf{k}\gamma_\mathbf{k}}{\omega_e^2}
\end{equation}
while, after keeping on the r.h.s. of the same equation only the electron terms, we get
\begin{equation}\label{eqthapp:uh gammak general}
\gamma_\mathbf{k} \simeq -\frac{\pi \omega_e^4}{|\Omega_e|^3}\frac{e^{-\xi^e_\mathbf{k}}}{\xi^e_\mathbf{k}}\sum \limits_{n=1}^\infty nI_n(\xi^e_\mathbf{k})\delta \left( n-\frac{\omega_k}{|\Omega_e|} \right) .
\end{equation}
Since $\omega_\mathbf{k}\simeq |\Omega_e|$ we have $|n|=1$, and if we use the relation
$$e^{-\xi^e_\mathbf{k}}\sum \limits_{n=1}^\infty \frac{nI_n(\xi^e_\mathbf{k})}{\xi^e_\mathbf{k}}\simeq [(\xi^e_\mathbf{k})^{-1}-1](\xi^e_\mathbf{k}/2) \simeq \frac{1}{2} \; , \; \; \text{for} \; \; \xi^e_\mathbf{k} \ll1,$$
we obtain the further simplified relation
\begin{equation}\label{eqapp:imaginary freq part solution}
\gamma_\mathbf{k} \simeq -\frac{\pi}{4}\frac{\omega_e^4}{|\Omega_e|^3}.
\end{equation}

For the case of LH waves, we search for solutions to Eq.~(\ref{eqapp:real freq equation}) with $\Omega_i \lesssim \omega_\mathbf{k}\ll |\Omega_e|$, and we find 
\begin{equation}\label{app:lh real freq part}
\omega_\mathbf{k}^2 \simeq \Omega_i^2+\frac{\omega_i^2}{1+(\omega_e/\Omega_e)^2} \simeq \Omega_i^2,
\end{equation}
if $\omega_i\ll \Omega_i$, and in the same way as in the UH case, 
for the imaginary part we find the solution
\begin{equation}\label{eqapp:lh imaginary part solution}
\gamma_\mathbf{k} \simeq -\frac{\pi}{4}\frac{\omega_i^4}{\Omega_i^3},
\end{equation}
since the electron contribution on the r.h.s.\ of Eq.~(\ref{eqapp:general dispersion relation}) vanishes due to the condition $\omega_e\ll |\Omega_e|$, which also holds in this case.

\section{Diffusion coefficient for Upper Hybrid and Lower Hybrid waves}
\label{sec:appendixB-diffusion coefficients}

Starting with Eq.~(\ref{eqth:velocity diff vector}), the vector $\mathbf{d}^s_{n,\mathbf{k}}$ is easily reduced to the expression (\ref{eqhybrid:diff vector}), where the recurrence relation $J_{n-1}(z)+J_{n+1}(z)=(2n/z)J_n(z)$ has been taken into account and we defined $\zeta ^s _\mathbf{k} \equiv k_\perp v_\perp /\Omega _s$. Then, the only non-zero component of the diffusion tensor (\ref{eqth:velocity diff tensor}) is the $\hat{\mathbf{e}}_\perp \hat{\mathbf{e}}_\perp$-term, which becomes
\begin{eqnarray}\label{eqapp:almost hybrid diffusion coefficient}
D ^{\rm QL} _{\perp s} & \simeq & \frac{8\pi e^2}{m_e^2} \sum \limits _{n=-\infty}^{\infty} \int d^3\mathbf{k} \; \frac{i\mathcal{W}_\mathbf{k} J_n^2(\zeta ^s _\mathbf{k})}{(\omega _\mathbf{k} -n\Omega _s)+i\gamma _\mathbf{k}}\left( \frac{n}{\zeta ^s _\mathbf{k}} \right) ^2 \nonumber \\	
&=& \frac{8\pi e^2}{m_e^2} \int d^3\mathbf{k} \; \mathcal{W}_\mathbf{k} \sum \limits _{n=-\infty}^ \infty n^2 \left[ \frac{J_n(\zeta ^s _\mathbf{k})}{\zeta^s_\mathbf{k}} \right] ^2 \nonumber \\
&& \times \left[ \mathcal{P} \frac{i(\omega_\mathbf{k} -n\Omega _s)+\gamma_\mathbf{k}}{(\omega _\mathbf{k} -n\Omega _s)^2+\gamma _\mathbf{k}^2}+\pi\delta \left( \omega_\mathbf{k}-n \Omega _s \right) \right] .
\end{eqnarray}

The wave frequency is a harmonic of the gyrofrequency, as the resonance condition requires, and in the limit $\omega_\mathbf{k}\rightarrow n\Omega_s$, Eq.~(\ref{eqapp:almost hybrid diffusion coefficient}) is simplified to
\begin{eqnarray}\label{eqapp:hybrid diffusion coefficient}
D ^{\rm QL} _{\perp s} \simeq &&  \frac{2\pi \omega_s^2}{n_0m_s}\sum \limits _{n=-\infty}^ \infty n^2 \nonumber \\
&& \times \int d^3\mathbf{k} \; \mathcal{W}_\mathbf{k} \left[ \frac{J_n(\zeta ^s _\mathbf{k})}{\zeta ^s _\mathbf{k}} \right] ^2 \delta \left( \omega_\mathbf{k} - n\Omega _s \right) ,
\end{eqnarray}
Thus, since in both applications, to UH and LH waves, respectively, the resonance includes only $|n|=1$, by using the approximation \cite{newman84}
$$\frac{J_1(\zeta^s_\mathbf{k})}{\zeta^s_\mathbf{k}}\simeq 0.500 -0.562 (\zeta^s_\mathbf{k}/3)^2+\cdots \simeq \frac{1}{2}\; , \; \text{for} \; |\zeta^s_\mathbf{k}| \ll 1,$$
we can further simplify the diffusion coefficients in Eq.~(\ref{eqapp:hybrid diffusion coefficient}) to the ones in Eq.~(\ref{eqthUH:diff coeff}).

\end{appendix}

\bibliography{ElectrostaticQLrefs}

\providecommand{\noopsort}[1]{}\providecommand{\singleletter}[1]{#1}%
\begin{thebibliography}{23}%
\makeatletter
\providecommand \@ifxundefined [1]{%
 \@ifx{#1\undefined}
}%
\providecommand \@ifnum [1]{%
 \ifnum #1\expandafter \@firstoftwo
 \else \expandafter \@secondoftwo
 \fi
}%
\providecommand \@ifx [1]{%
 \ifx #1\expandafter \@firstoftwo
 \else \expandafter \@secondoftwo
 \fi
}%
\providecommand \natexlab [1]{#1}%
\providecommand \enquote  [1]{``#1''}%
\providecommand \bibnamefont  [1]{#1}%
\providecommand \bibfnamefont [1]{#1}%
\providecommand \citenamefont [1]{#1}%
\providecommand \href@noop [0]{\@secondoftwo}%
\providecommand \href [0]{\begingroup \@sanitize@url \@href}%
\providecommand \@href[1]{\@@startlink{#1}\@@href}%
\providecommand \@@href[1]{\endgroup#1\@@endlink}%
\providecommand \@sanitize@url [0]{\catcode `\\12\catcode `\$12\catcode
  `\&12\catcode `\#12\catcode `\^12\catcode `\_12\catcode `\%12\relax}%
\providecommand \@@startlink[1]{}%
\providecommand \@@endlink[0]{}%
\providecommand \url  [0]{\begingroup\@sanitize@url \@url }%
\providecommand \@url [1]{\endgroup\@href {#1}{\urlprefix }}%
\providecommand \urlprefix  [0]{URL }%
\providecommand \Eprint [0]{\href }%
\providecommand \doibase [0]{http://dx.doi.org/}%
\providecommand \selectlanguage [0]{\@gobble}%
\providecommand \bibinfo  [0]{\@secondoftwo}%
\providecommand \bibfield  [0]{\@secondoftwo}%
\providecommand \translation [1]{[#1]}%
\providecommand \BibitemOpen [0]{}%
\providecommand \bibitemStop [0]{}%
\providecommand \bibitemNoStop [0]{.\EOS\space}%
\providecommand \EOS [0]{\spacefactor3000\relax}%
\providecommand \BibitemShut  [1]{\csname bibitem#1\endcsname}%
\let\auto@bib@innerbib\@empty
\bibitem [{\citenamefont {Kennel}\ and\ \citenamefont
  {Englemann}(1966)}]{AIPformalism}%
  \BibitemOpen
  \bibfield  {author} {\bibinfo {author} {\bibfnamefont {C.~F.}\ \bibnamefont
  {Kennel}}\ and\ \bibinfo {author} {\bibfnamefont {F.}~\bibnamefont
  {Englemann}},\ }\href@noop {} {\bibfield  {journal} {\bibinfo  {journal}
  {AIP, Phys. Fluids}\ }\textbf {\bibinfo {volume} {9}},\ \bibinfo {pages}
  {2377} (\bibinfo {year} {1966})}\BibitemShut {NoStop}%
\bibitem [{\citenamefont {Lerche}(1968)}]{lerche68}%
  \BibitemOpen
  \bibfield  {author} {\bibinfo {author} {\bibfnamefont {I.}~\bibnamefont
  {Lerche}},\ }\href@noop {} {\bibfield  {journal} {\bibinfo  {journal} {Phys.
  Fluids}\ }\textbf {\bibinfo {volume} {11}},\ \bibinfo {pages} {1720}
  (\bibinfo {year} {1968})}\BibitemShut {NoStop}%
\bibitem [{\citenamefont {Tao}\ \emph {et~al.}(2011)\citenamefont {Tao},
  \citenamefont {Bortnik}, \citenamefont {Albert}, \citenamefont {Liu},\ and\
  \citenamefont {Thorne}}]{tao11}%
  \BibitemOpen
  \bibfield  {author} {\bibinfo {author} {\bibfnamefont {X.}~\bibnamefont
  {Tao}}, \bibinfo {author} {\bibfnamefont {J.}~\bibnamefont {Bortnik}},
  \bibinfo {author} {\bibfnamefont {J.~M.}\ \bibnamefont {Albert}}, \bibinfo
  {author} {\bibfnamefont {K.}~\bibnamefont {Liu}}, \ and\ \bibinfo {author}
  {\bibfnamefont {R.~M.}\ \bibnamefont {Thorne}},\ }\href@noop {} {\bibfield
  {journal} {\bibinfo  {journal} {Geophys. Res. Lett.}\ }\textbf {\bibinfo
  {volume} {38}},\ \bibinfo {pages} {L06105} (\bibinfo {year}
  {2011})}\BibitemShut {NoStop}%
\bibitem [{\citenamefont {Karney}(1979)}]{karney77II}%
  \BibitemOpen
  \bibfield  {author} {\bibinfo {author} {\bibfnamefont {C.~F.~F.}\
  \bibnamefont {Karney}},\ }\href@noop {} {\bibfield  {journal} {\bibinfo
  {journal} {Phys. Fluids}\ }\textbf {\bibinfo {volume} {22}},\ \bibinfo
  {pages} {2188} (\bibinfo {year} {1979})}\BibitemShut {NoStop}%
\bibitem [{\citenamefont {Karney}(1978)}]{karneyAIP78}%
  \BibitemOpen
  \bibfield  {author} {\bibinfo {author} {\bibfnamefont {C.~F.~F.}\
  \bibnamefont {Karney}},\ }\href@noop {} {\bibfield  {journal} {\bibinfo
  {journal} {Phys. Fluids}\ }\textbf {\bibinfo {volume} {21}},\ \bibinfo
  {pages} {1584} (\bibinfo {year} {1978})}\BibitemShut {NoStop}%
\bibitem [{\citenamefont {Smith}\ and\ \citenamefont
  {Kaufman}(1978)}]{kaufman79}%
  \BibitemOpen
  \bibfield  {author} {\bibinfo {author} {\bibfnamefont {G.~R.}\ \bibnamefont
  {Smith}}\ and\ \bibinfo {author} {\bibfnamefont {A.~N.}\ \bibnamefont
  {Kaufman}},\ }\href@noop {} {\bibfield  {journal} {\bibinfo  {journal} {Phys.
  Fluids}\ }\textbf {\bibinfo {volume} {21}},\ \bibinfo {pages} {2230}
  (\bibinfo {year} {1978})}\BibitemShut {NoStop}%
\bibitem [{\citenamefont {Gell}\ and\ \citenamefont {Nakach}(1980)}]{gell80}%
  \BibitemOpen
  \bibfield  {author} {\bibinfo {author} {\bibfnamefont {Y.}~\bibnamefont
  {Gell}}\ and\ \bibinfo {author} {\bibfnamefont {R.}~\bibnamefont {Nakach}},\
  }\href@noop {} {\bibfield  {journal} {\bibinfo  {journal} {Phys. Fluids}\
  }\textbf {\bibinfo {volume} {23}},\ \bibinfo {pages} {1646} (\bibinfo {year}
  {1980})}\BibitemShut {NoStop}%
\bibitem [{\citenamefont {Shklyar}(1986)}]{shklyar86}%
  \BibitemOpen
  \bibfield  {author} {\bibinfo {author} {\bibfnamefont {D.~R.}\ \bibnamefont
  {Shklyar}},\ }\href@noop {} {\bibfield  {journal} {\bibinfo  {journal}
  {Planet. Space Sci.}\ }\textbf {\bibinfo {volume} {34}},\ \bibinfo {pages}
  {1091} (\bibinfo {year} {1986})}\BibitemShut {NoStop}%
\bibitem [{\citenamefont {Benkadda}, \citenamefont {Sen},\ and\ \citenamefont
  {Shklyar}(1996)}]{sen96}%
  \BibitemOpen
  \bibfield  {author} {\bibinfo {author} {\bibfnamefont {S.}~\bibnamefont
  {Benkadda}}, \bibinfo {author} {\bibfnamefont {A.}~\bibnamefont {Sen}}, \
  and\ \bibinfo {author} {\bibfnamefont {D.~R.}\ \bibnamefont {Shklyar}},\
  }\href@noop {} {\bibfield  {journal} {\bibinfo  {journal} {American Institute
  of Physics, CHAOS}\ }\textbf {\bibinfo {volume} {6}},\ \bibinfo {pages} {451}
  (\bibinfo {year} {1996})}\BibitemShut {NoStop}%
\bibitem [{\citenamefont {Lange}\ \emph {et~al.}(2013)\citenamefont {Lange},
  \citenamefont {Spanier}, \citenamefont {Battarbee}, \citenamefont {Vainio},\
  and\ \citenamefont {Laitinen}}]{lange13}%
  \BibitemOpen
  \bibfield  {author} {\bibinfo {author} {\bibfnamefont {S.}~\bibnamefont
  {Lange}}, \bibinfo {author} {\bibfnamefont {F.}~\bibnamefont {Spanier}},
  \bibinfo {author} {\bibfnamefont {M.}~\bibnamefont {Battarbee}}, \bibinfo
  {author} {\bibfnamefont {R.}~\bibnamefont {Vainio}}, \ and\ \bibinfo {author}
  {\bibfnamefont {T.}~\bibnamefont {Laitinen}},\ }\href@noop {} {\bibfield
  {journal} {\bibinfo  {journal} {Astronomy \& Astrophysics}\ }\textbf
  {\bibinfo {volume} {553}},\ \bibinfo {pages} {A129} (\bibinfo {year}
  {2013})}\BibitemShut {NoStop}%
\bibitem [{\citenamefont {Shalchi}\ \emph {et~al.}(2004)\citenamefont
  {Shalchi}, \citenamefont {Bieber}, \citenamefont {Matthaeus},\ and\
  \citenamefont {Qin}}]{shalchi04}%
  \BibitemOpen
  \bibfield  {author} {\bibinfo {author} {\bibfnamefont {A.}~\bibnamefont
  {Shalchi}}, \bibinfo {author} {\bibfnamefont {J.~W.}\ \bibnamefont {Bieber}},
  \bibinfo {author} {\bibfnamefont {W.~H.}\ \bibnamefont {Matthaeus}}, \ and\
  \bibinfo {author} {\bibfnamefont {G.}~\bibnamefont {Qin}},\ }\href@noop {}
  {\bibfield  {journal} {\bibinfo  {journal} {The Astrophysical Journal}\
  }\textbf {\bibinfo {volume} {616}},\ \bibinfo {pages} {617} (\bibinfo {year}
  {2004})}\BibitemShut {NoStop}%
\bibitem [{\citenamefont {Lange}\ and\ \citenamefont
  {Spanier}(2012)}]{spanier12}%
  \BibitemOpen
  \bibfield  {author} {\bibinfo {author} {\bibfnamefont {S.}~\bibnamefont
  {Lange}}\ and\ \bibinfo {author} {\bibfnamefont {F.}~\bibnamefont
  {Spanier}},\ }\href@noop {} {\bibfield  {journal} {\bibinfo  {journal}
  {Astronomy \& Astrophysics}\ }\textbf {\bibinfo {volume} {546}},\ \bibinfo
  {pages} {A51} (\bibinfo {year} {2012})}\BibitemShut {NoStop}%
\bibitem [{\citenamefont {Sagdeev}\ and\ \citenamefont
  {Galeev}(1969)}]{galeev69}%
  \BibitemOpen
  \bibfield  {author} {\bibinfo {author} {\bibfnamefont {R.~Z.}\ \bibnamefont
  {Sagdeev}}\ and\ \bibinfo {author} {\bibfnamefont {A.~A.}\ \bibnamefont
  {Galeev}},\ }\href@noop {} {\emph {\bibinfo {title} {Nonlinear plasma
  theory}}}\ (\bibinfo  {publisher} {W. A. Benjamin, Inc.},\ \bibinfo {year}
  {1969})\BibitemShut {NoStop}%
\bibitem [{\citenamefont {Nicholson}(1983)}]{dwight93}%
  \BibitemOpen
  \bibfield  {author} {\bibinfo {author} {\bibfnamefont {D.~R.}\ \bibnamefont
  {Nicholson}},\ }\href@noop {} {\emph {\bibinfo {title} {Introduction to
  plasma theory}}}\ (\bibinfo  {publisher} {John Wiley \& Sons, Inc.},\
  \bibinfo {year} {1983})\BibitemShut {NoStop}%
\bibitem [{\citenamefont {Frisch}(1995)}]{kolm95}%
  \BibitemOpen
  \bibfield  {author} {\bibinfo {author} {\bibfnamefont {U.}~\bibnamefont
  {Frisch}},\ }\href@noop {} {\emph {\bibinfo {title} {Turbulence: the legacy
  of A. N. Kolmogorov}}}\ (\bibinfo  {publisher} {Cambridge University Press},\
  \bibinfo {year} {1995})\BibitemShut {NoStop}%
\bibitem [{\citenamefont {Chirikov}(1979)}]{chirikov79}%
  \BibitemOpen
  \bibfield  {author} {\bibinfo {author} {\bibfnamefont {B.~V.}\ \bibnamefont
  {Chirikov}},\ }\href@noop {} {\bibfield  {journal} {\bibinfo  {journal}
  {Physics Reports}\ }\textbf {\bibinfo {volume} {52}},\ \bibinfo {pages} {263}
  (\bibinfo {year} {1979})}\BibitemShut {NoStop}%
\bibitem [{\citenamefont {Diamond}, \citenamefont {Itoh},\ and\ \citenamefont
  {Itoh}(2009)}]{itoh09}%
  \BibitemOpen
  \bibfield  {author} {\bibinfo {author} {\bibfnamefont {P.~H.}\ \bibnamefont
  {Diamond}}, \bibinfo {author} {\bibfnamefont {S.-I.}\ \bibnamefont {Itoh}}, \
  and\ \bibinfo {author} {\bibfnamefont {K.}~\bibnamefont {Itoh}},\ }\href@noop
  {} {\emph {\bibinfo {title} {Modern Plasma Physics Vol. 1: Physical Kinetics
  of Turbulent Plasmas}}}\ (\bibinfo  {publisher} {Cambridge University
  Press},\ \bibinfo {year} {2009})\BibitemShut {NoStop}%
\bibitem [{\citenamefont {Shapiro}\ and\ \citenamefont
  {Sagdeev}(1997)}]{shapiro97}%
  \BibitemOpen
  \bibfield  {author} {\bibinfo {author} {\bibfnamefont {V.~D.}\ \bibnamefont
  {Shapiro}}\ and\ \bibinfo {author} {\bibfnamefont {R.~Z.}\ \bibnamefont
  {Sagdeev}},\ }\href@noop {} {\bibfield  {journal} {\bibinfo  {journal}
  {Physics Reports}\ }\textbf {\bibinfo {volume} {283}},\ \bibinfo {pages} {49}
  (\bibinfo {year} {1997})}\BibitemShut {NoStop}%
\bibitem [{\citenamefont {Chew}, \citenamefont {Goldberger},\ and\
  \citenamefont {Low}(1956)}]{chew56}%
  \BibitemOpen
  \bibfield  {author} {\bibinfo {author} {\bibfnamefont {G.~F.}\ \bibnamefont
  {Chew}}, \bibinfo {author} {\bibfnamefont {M.~L.}\ \bibnamefont
  {Goldberger}}, \ and\ \bibinfo {author} {\bibfnamefont {F.~E.}\ \bibnamefont
  {Low}},\ }\href@noop {} {\bibfield  {journal} {\bibinfo  {journal} {Proc.
  Loy. Soc.}\ }\textbf {\bibinfo {volume} {236}},\ \bibinfo {pages} {112}
  (\bibinfo {year} {1956})}\BibitemShut {NoStop}%
\bibitem [{\citenamefont {Bittencourt}(2004)}]{bitten04}%
  \BibitemOpen
  \bibfield  {author} {\bibinfo {author} {\bibfnamefont {L.~A.}\ \bibnamefont
  {Bittencourt}},\ }\href@noop {} {\emph {\bibinfo {title} {Fundamentals of
  plasma physics}}}\ (\bibinfo  {publisher} {Springer-Verlag New York},\
  \bibinfo {year} {2004})\BibitemShut {NoStop}%
\bibitem [{\citenamefont {Ragwitz}\ and\ \citenamefont
  {Kantz}(2001)}]{Ragwitz2001}%
  \BibitemOpen
  \bibfield  {author} {\bibinfo {author} {\bibfnamefont {M.}~\bibnamefont
  {Ragwitz}}\ and\ \bibinfo {author} {\bibfnamefont {H.}~\bibnamefont
  {Kantz}},\ }\href@noop {} {\bibfield  {journal} {\bibinfo  {journal}
  {Physical Review Letters}\ }\textbf {\bibinfo {volume} {87}},\ \bibinfo
  {pages} {254501} (\bibinfo {year} {2001})}\BibitemShut {NoStop}%
\bibitem [{\citenamefont {Abramowitz}\ and\ \citenamefont
  {(Editors)}(1972)}]{abra72}%
  \BibitemOpen
  \bibfield  {author} {\bibinfo {author} {\bibfnamefont {M.}~\bibnamefont
  {Abramowitz}}\ and\ \bibinfo {author} {\bibfnamefont {I.~A.~S.}\ \bibnamefont
  {(Editors)}},\ }\href@noop {} {\emph {\bibinfo {title} {Handbook of
  Mathematical functions}}}\ (\bibinfo  {publisher} {National Bureau of
  Standards, Applied Mathematics Series},\ \bibinfo {year} {1972})\BibitemShut
  {NoStop}%
\bibitem [{\citenamefont {Newman}(1984)}]{newman84}%
  \BibitemOpen
  \bibfield  {author} {\bibinfo {author} {\bibfnamefont {J.~N.}\ \bibnamefont
  {Newman}},\ }\href@noop {} {\emph {\bibinfo {title} {Approximations for the
  Bessel and Struve functions}}}\ (\bibinfo  {publisher} {American Mathematical
  Society, Mathematics of computation},\ \bibinfo {year} {1984})\BibitemShut
  {NoStop}%
\end{thebibliography}%

\end{document}